\documentclass[sigconf]{acmart}
\AtBeginDocument{%
  }
\setcopyright{acmlicensed}

\setcopyright{none}
\renewcommand\footnotetextcopyrightpermission[1]{}
\acmYear{2025}
\copyrightyear{2025}
\acmConference[SoCC '25]{ACM Symposium on Cloud Computing}{November 19--21, 2025}{Online, }
\acmBooktitle{ACM Symposium on Cloud Computing (SoCC '25), November 19--21, 2025, Online, }
\acmDOI{10.1145/3772052.3772230}
\acmISBN{979-8-4007-2276-9/25/11}

\newcommand{\scheme}{Symbiosis}

\usepackage{tikz}
\usepackage{amsmath}
\usepackage{lipsum}
\usepackage{subcaption}
\usepackage{enumitem}
\usepackage{caption}
\usepackage{array, booktabs}
\newcolumntype{P}[1]{>{\centering\arraybackslash}p{#1}}
\usepackage{xcolor, colortbl}
\usepackage[utf8]{inputenc}
\usepackage[font=small]{caption}
\usepackage{bbm}
\begin{document}

\title{\scheme: Multi-Adapter Inference and Fine-Tuning}

\author{
{\rm Saransh Gupta}
\footnotemark[1]
\qquad
{\rm Umesh Deshpande}
\footnotemark[1]
\qquad
{\rm Travis Janssen}
\qquad
{\rm Swaminathan Sundararaman}
}
\affiliation{%
    \institution{IBM Research, USA}
    \country{}
}

\begin{abstract}
Parameter-efficient fine-tuning (PEFT) allows model builders to capture the task-specific parameters into adapters, which are a fraction of the size of the original base model. Popularity of PEFT technique for fine-tuning has led to the creation of a large number of adapters for popular Large Language Models (LLMs). However, existing frameworks fall short in supporting inference or fine-tuning with multiple adapters in the following ways. 1) For fine-tuning, each job needs to deploy its dedicated base model instance, which results in excessive GPU memory consumption and poor GPU utilization. 2) While popular inference platforms can serve multiple PEFT adapters, they do not allow independent resource management or mixing of different PEFT methods. 3) They cannot make effective use of heterogeneous accelerators. 4) They do not provide privacy to users who may not wish to expose their fine-tuned parameters to service providers. In \scheme{}, we address the above problems by enabling the as-a-service deployment of the base model. The base model layers can be shared across multiple inference or fine-tuning processes. Our split-execution technique decouples the execution of client-specific adapters and layers from the frozen base model layers offering them flexibility to manage their resources, to select their fine-tuning method, to achieve their performance goals. Our approach is transparent to models and works out-of-the-box for most models in the transformers library. We demonstrate the use of \scheme{} to simultaneously fine-tune 20 Gemma2-27B LoRA adapters on 8 GPUs.
\end{abstract}

\maketitle

\footnotetext[1]{
Corresponding Authors: saransh@ibm.com, udeshpa@us.ibm.com
}

\section{Introduction}
\label{sec:intro}
Parameter Efficient Fine Tuning (PEFT)~\cite{peft} is a popular method to fine-tune the existing foundation models with various fine-tuning methods, e.g., LoRA, IA3, and AdaLoRA. The adapters created using PEFT are a fraction of the size of the base model used, often consuming only 10s of MBs of accelerator memory, as opposed to the base model that consumes several Gigabytes. The resulting efficiency and low cost in fine-tuning adapters have led to the creation of hundreds of adapters for many popular models, such as Llama, Gemma, StarCoder, etc. For instance, an LLM agentic system may fine-tune and serve several adapters that address different use cases. Or, they may want to evaluate a range of adapter hyperparameters, e.g., LoRA rank, alpha, etc. to reach the best possible model accuracy.

\begin{figure}[h]
    \centering
    \includegraphics[width=0.95\linewidth]{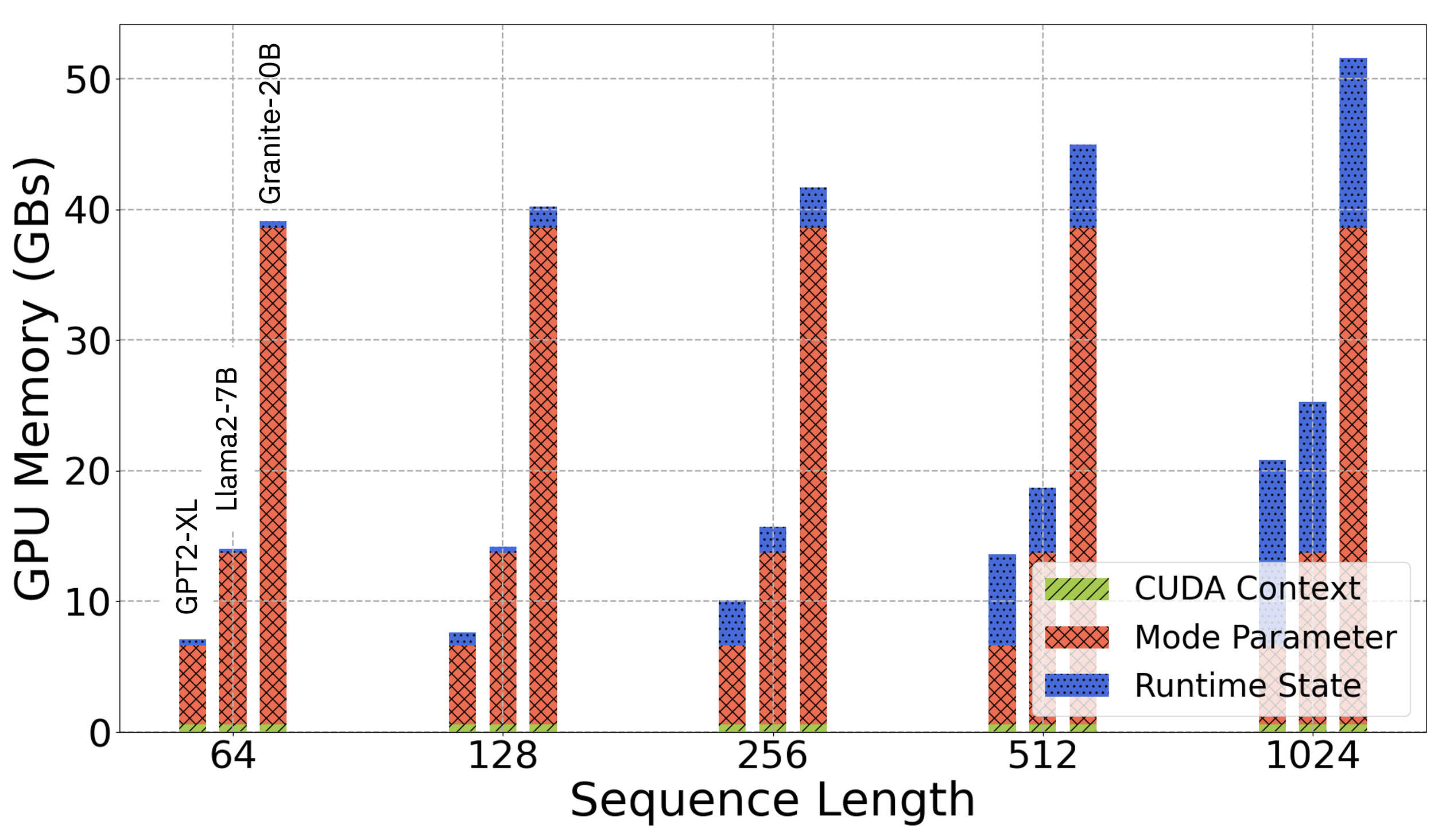}
    \caption{GPU memory consumption for fine-tuning of a single rank-8 LoRA adapter for GPT2-XL, Llama2-7B, and Granite-20B. Runtime state requires GBs of GPU memory, especially at larger sequence lengths.}
    \label{fig:memory-motivation}
\end{figure}

However, the existing platforms~\cite{transformers} that support simultaneous fine-tuning of multiple PEFT adapters tend to underutilize GPU's computational capability~\cite{splitwise, sheng_s-lora_2024}. Users must launch separate jobs for different adapters, each requiring a dedicated model instance. The model instances occupy GPU memory while not fully exploiting its computational capability, especially when the jobs do not receive enough requests.
Recent works~\cite{punica, sheng_s-lora_2024, mLoRA2023, flexllm} address the problem by sharing the model instance across inference or fine-tuning jobs to reduce its memory footprint and increase the computational density. However, when multiple jobs share the same GPUs, their runtime state (e.g., KV cache, optimizer state) competes with model instance and other jobs for GPU resources (Figure~\ref{fig:memory-motivation}). In such systems, sharing GPUs across different jobs entails the burden of having to deal with their dynamic GPU memory demands from either varying rate of requests or context lengths.


\begin{figure}[h]
    \centering
    \includegraphics[width=\linewidth]{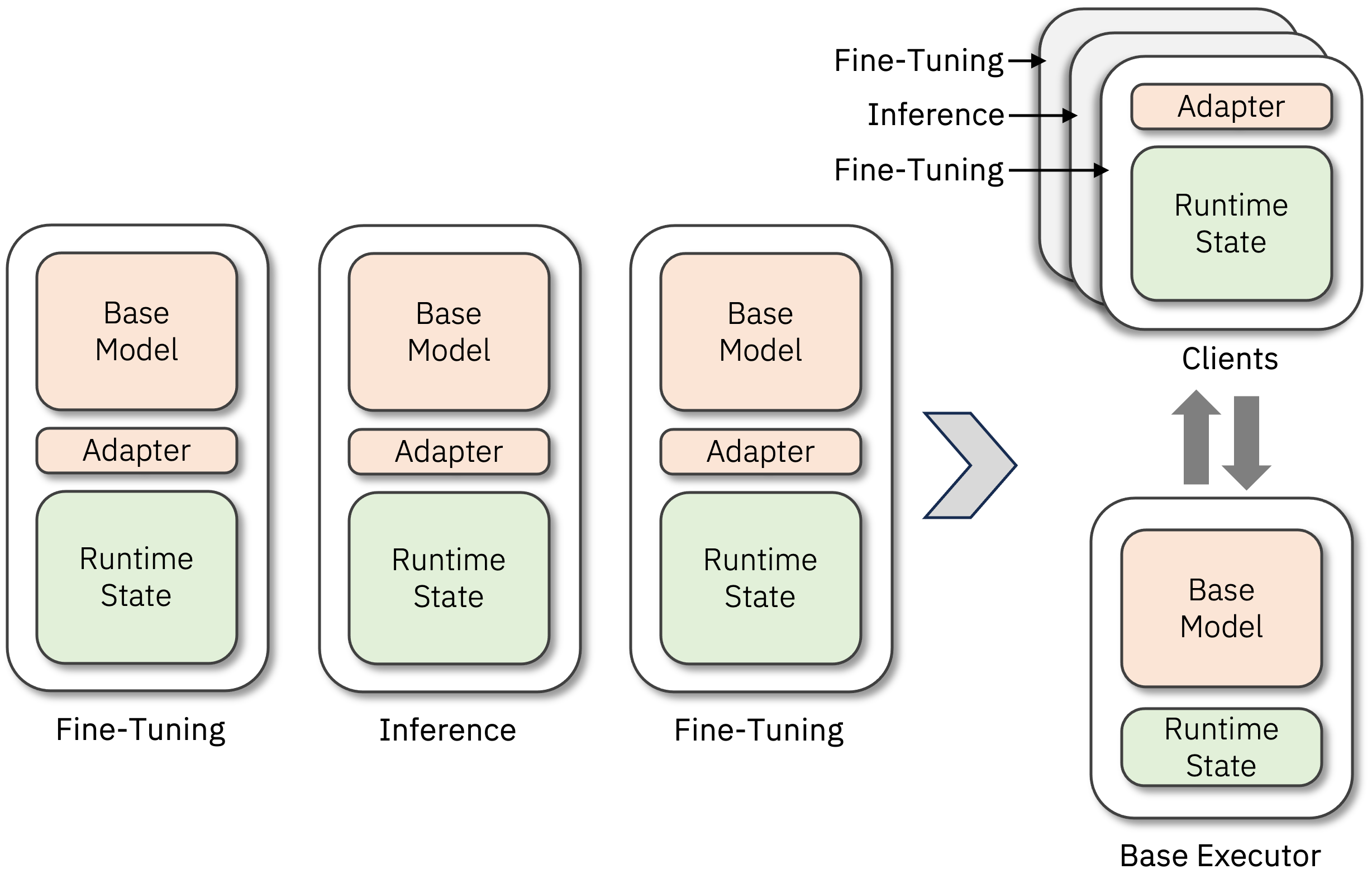}
    \caption{GPU memory layout with baseline (left) and \scheme{} (right).}
    \label{fig:split-exec}
\end{figure}

We address the above problems in {\em \scheme{}}. \scheme{} shares the common model parameters across the clients for inference and/or fine-tuning jobs. The shared fraction of the model is referred to as {\em base model}.
We transparently separate the execution of the base model (in {\em base executor}) from that of the job specific components, namely, attention and adapter layers, e.g., LoRA, (in {\em client}). This {\em split execution} offers the following benefits.
(1) The separation of execution allows a single base executor to serve multiple inference or fine-tuning clients, where the base executor can be placed in one or more GPUs independently from its clients. This means the base executor can span across multiple GPUs and the clients can be either co-located with the base executor, placed on different GPUs, or even placed on a different node. This is important because the clients' resource consumption pattern differs from that of the base executor. The client contains request's runtime state, such as, KV cache or optimizer state, which grows linearly with each new generated token (for inference) or each new pass (for fine-tuning). In contrast, the base executor is stateless. Its resource consumption directly reflects the number of tokens being actively processed at any given moment. Therefore, by separating the base executor from the client, the client can be placed independently and remains insulated from runtime state fluctuations caused by other clients.
(2) The clients are computationally less intensive, thus they can be hosted on less powerful GPUs or CPUs with little performance degradation.
(3) Each user can independently configure its client for an inference or fine-tuning job with a preferred PEFT method while sharing the base model instance. Users opt for different PEFT methods for their different use-case specific strengths, inference overheads, and fine-tuning resource requirements~\cite{peft-comparison, ia3, rosa, peft-compare}. (4) Different clients using the common base executor need not be in lockstep for the execution of base model layers. Each client can drive its execution at a different rate, while opportunistically sharing the execution of the base model with other clients at a layer granularity.

The technical contributions of \scheme{} are as follows.
\begin{itemize}
    \item \scheme{} proposes a general purpose framework to share base model(s) across multiple inference and fine-tuning jobs located across GPUs on same or different nodes. 
    \item \scheme{} presents a split-execution technique, which 
    identifies the base model and client-specific parts of the model structure and transparently decouples them.
    This means \scheme{} works for a variety of model architectures (e.g., Llama, GPT) and a variety of PEFT methods (e.g., LoRA, IA3) without requiring any changes to model code. We demonstrate this with 4 different model architectures.
    \item \scheme{} presents an opportunistic batching technique to batch the inference and fine-tuning requests from different clients at the base executor, thereby improving computational efficiency. Even when the requests are of different token lengths, base executor can batch them together without requiring any padding for shorter requests, avoiding the computational overhead of padding.
    \item \scheme{} presents a technique to preserve the privacy of model activations communicated across the clients and base executor. This enables the privacy sensitive clients to use an un-trusted base model service provider without having to expose the activations.
\end{itemize}



We demonstrate the following use-cases in \scheme{}.

\paragraph{Base Model as-a-service:}
\scheme{} enables the cloud providers to serve a base model instance to multiple users for inference and fine-tuning jobs. This use case achieves the following goals. First, it reduces the model serving cost for users by reducing their burden of hosting an entire model. The users can share the common base model with other users. Second, it simplifies the service provider's resource management. It offloads the responsibility of fulfilling the job specific resource demand to users. For instance, a request's context length or an optimizer selected for a fine-tuning job determines the client's GPU memory requirement. The service providers are isolated from such requirements. They only need to provision the base executor resources to fulfill the expected token processing rate for a given base model, which is made easier because the per-token resource requirement remains constant irrespective of the client-side configurations.

\paragraph{Use of Heterogeneous Accelerators:}
Hardware upgrade with the release of new GPUs is often staggered, which results in nodes in a cluster containing different generations of GPUs. Lack of consideration for their different capabilities degrades the workload performance. \scheme{} proposes a better way to leverage their different capabilities. It decouples the execution of computationally heavy base model from the comparatively lighter clients. This allows the base model to always use more powerful devices with the clients being potentially placed on less powerful devices, without significant impact on inference or fine-tuning performance. Moreover, \scheme{} also enables execution of clients on CPUs. This is particularly useful for jobs with large KV cache (e.g., with longer sequence lengths) that require 100s of GBs of memory. Longer contexts are common in Retrieval Augmented Generation (RAG)~\cite{rag}, where an additional context is retrieved from a knowledge source and appended to the original prompt. Leveraging the larger CPU memory for such jobs reduces fine-tuning and inference cost. We demonstrate in the evaluation section that for 64K context length, \scheme{} performs CPU-GPU inference and achieves 33\% speed up compared to the GPU-only baseline.

\paragraph{Privacy Preserving Multi-Tenant Platform:}
With \scheme{}, multiple customers can safely leverage a third-party base model service to fine-tune their models. Even though the popular base model parameters are publicly available, customers may not wish to expose their adapter parameters (which may be trained on customers' confidential data) and activations to a third-party base model service provider. The challenge of separating the adapter parameters is naturally addressed by \scheme{}'s decoupled execution. This allows a tenant to host the tenant-specific computation in a secure environment. Moreover, \scheme{} also provides a mechanism that avoids exposing the activations that are communicated across client and base executor in order to protect the adapter parameters against model extraction attacks~\cite{me-attack, honeypotnet, me-attack2}. These attacks observe the model activations to infer the model parameters.
\section{Related Work}
\subsection{Model Sharing in Fine Tuning and Inference}
\paragraph{Fine-tuning}
Most commonly used PEFT fine-tuning systems (like PyTorch, HF-Trainer \cite{tx}) target single adapter fine-tuning where multi-adapter fine-tuning can only be achieved via multiple isolated tuning jobs.
MixLoRA~\cite{li2024mixlora} supports fine-tuning of multiple LoRA adapters for multiple experts in Mixture of Expert model. In contrast, \scheme{} proposes a general purpose system.
FlexLLM~\cite{flexllm} enables sharing of the base model across multiple fine-tuning jobs. However, it relies on static compilation of models~\cite{flexflow} to generate execution graphs and parallelization strategies based on the model's architecture. Therefore, any structural changes to model architecture, such as adding new PEFT adapters, cannot occur at runtime. In contrast, \scheme{} is designed for Transformers platform to handle the addition (or removal) of clients (i.e., PEFT adapters) at runtime. This gives tenants runtime control over their clients' placement, thus enabling better resource isolation and privacy.
mLoRA~\cite{aspen, mLoRA2023} can simultaneously fine-tune multiple LoRA adapters by sharing the base model across the trainers. However, it only supports variants of LoRA fine-tuning methods. Whereas, \scheme{} allows fine-tuning of wider variety PEFT methods, e.g., IA3, P-tuning, Prefix-tuning etc. Second, unlike mLoRA, \scheme{} does not need to modify the model or user code. Therefore, \scheme{} can work with all the supported models in HuggingFace Transformers library~\cite{tx} out-of-the-box.
Finally, a key difference between mLoRA and \scheme{} is that in \scheme{}, each client is independent and is a driver of its training or inference. Because of this design principle each client can independently decide its training speed or select different PEFT fine-tuning method.

\paragraph{Inference} 
For inference, vLLM~\cite{noauthor_vllm-projectvllm_2024, sheng_s-lora_2024}, Punica~\cite{punica}, LoRAX~\cite{lorax2024}, Caraserve~\cite{noauthor_caraserve_nodate} enable sharing of a base model across different adapters and reduce memory consumption. Moreover, their heterogeneous batching allows LoRA adapters with different characteristics, e.g., rank, to be batched together. dLoRA~\cite{dlora} dynamically switches between merged and un-merged adapter to reduce request latency.
However, during inference in these systems, different adapters execute as a part the same process as the base model. This means, the client-specific data, i.e., adapter parameters, KV cache and activations, cannot be shielded from being accessed by the base model service provider. Also, clients cannot isolate themselves from the changing resource demands (e.g., increasing KV cache) of other clients. Moreover, the above platforms execute multiple adapters in lock step. Which means the clients with shorter sequences or smaller batches need to wait for the execution of the clients with longer sequences or larger batches at each layer. \scheme{}'s split execution addresses these challenges by isolating the clients from each other and only opportunistically sharing the execution of base model layers.


\subsection{Offloading KV Cache and Attention}
Transformers~\cite{transformers}, LMDeploy~\cite{lmdeploy2023}, CachedAttention~\cite{cachedattention}, Symphony~\cite{symphony}, DeepSpeed~\cite{deepspeed, deepspeed2, zeroinfinity} offload KV caches to host memory and storage to accommodate more requests or to improve the efficiency of multi-turn conversations. Splitwise~\cite{splitwise}, Mooncake~\cite{mooncake}, DistServe~\cite{distserve} split the prefill and generation phases on different GPUs or machines and allows phase specific resource management. They transfer KV cache from the prefill machine to the generation machine as the requests enters generation. \scheme{} leverages the OffloadCache~\cite{offloadedcache} feature of Transformers to offload the KV cache to host memory after prefill.
\scheme{} not only uses the host memory to store very large KV cache but also uses CPUs to execute clients and perform heterogeneous CPU-GPU inference for token generation. Moreover, unlike \scheme{}, the goals of these systems are not to provide client control and isolation. In these approaches, the KV cache of multiple clients is handled in the same manner.


Attention offloading is also well researched in LLMs. The attention computation is memory bound as opposed to the linear layers that are compute bound, thus it can better leverage non-GPU accelerators. FlexGen~\cite{flexgen}, Lamina~\cite{lamina}, LayerKV~\cite{layerkv} offload attention computation to CPU while offloading the KV caches to host memory and storage. InstInfer~\cite{pan2024instinfer} offloads attention to computational storage and uses its internal bandwidth and computational capability to improve inference. However, the above approaches are not transparent to model architectures, thus are not generally applicable. 

Infinigen~\cite{infinigen}, H2O~\cite{h2o}, DuoAttention~\cite{duo-attention} perform partial attention by focusing on important KV entries, while offloading less important entries to CPU memory. This allows them to support longer contexts while reducing GPU memory requirement. While \scheme{} can also benefit from reduced attention computation on GPU, especially when the client is located on CPU, we do not alter the default attention logic of the model provided in Transformers library.




\subsection{Efficient Batching for Performance}
Efficient batching is well explored for LLMs~\cite{cellbatch, orca, splitwise}.
Orca~\cite{orca} proposes continuous batching for inference requests, which instead of waiting for all requests in a batch to complete, batches new requests along with the ongoing requests. This reduces wasted computation on differently sized requests and improves throughput. Several research works also leverage the different resource consumption and performance characteristics of prefill and generation phases. Specifically, the prefill phase can saturate GPUs at smaller batches, whereas the generation phase performs better at larger batches. Sarathi~\cite{sarathi} proposes chunked prefill to split the prefill requests into smaller chunks and the remaining slots are filled with generation phase requests. This leads to better utilization of GPUs and improves inference throughput. \scheme{} fundamentally differs from the above works in that its base model layers serve a mix of inference (prefill or generation) and fine-tuning requests. Because of their different execution speeds and performance goals, in \scheme{}, the requests batched for the execution of the first layer need not be batched together for the execution of the subsequent layers. Therefore, we employ per-layers batching policies that honor low latency and improve system throughput.
\section{Design and Implementation}
\label{sec:design}

\subsection{Design Goals}


\begin{enumerate}

\item {\bf Model Sharing:} \scheme{} should allow fine-tuning and inference jobs to share base model parameters.

\item {\bf Flexible Placement:} A client should be able to (a) share GPUs with the base executor (e.g., in a resource constrained environment), (b) execute on a different GPU (to avoid interference), (c) execute on a CPU (e.g, to accommodate large KV cache) or (d) execute on another node (e.g., for privacy).
This is particularly important to operate in an heterogeneous environment, consisting of different generations of GPUs. \scheme{} should be able to leverage such resources while minimizing their performance impact.

\item {\bf Model Transparency:} The technique needs to be portable to different models, so that developers need not modify the model code (e.g., in the transformer library). This makes \scheme{} readily usable for most current and future models.

\item {\bf Batched Inference and Fine-tuning:} \scheme{} should be able to batch inference and fine-tuning requests from different clients. This opens up an opportunity to improve GPU utilization.

 \item {\bf Client Independence:} 
The requests should be able to progress through the base model at different rates. This is because the clients with larger adapters (e.g., higher LoRA rank), longer sequence lengths, or located on slower GPUs or CPU require more time for an iteration, and need not be executed in lockstep with latency-sensitive requests.

\item {\bf Multiple PEFT Methods:} \scheme{} should support simultaneous inference and fine-tuning for a mix of different PEFT methods.
\item {\bf Adapter Privacy:} The tenants should be able to use a shared base model (e.g., deployed by service provider) without having to expose their activations and adapter-specific model parameters.
\end{enumerate}

\begin{table*}[t]
    \centering
    \begin{tabular}{P{1.8cm}P{1.2cm}P{1.2cm}P{1.2cm}P{1.5cm}P{1.4cm}P{1.4cm}P{1.4cm}P{1.4cm}P{1.2cm}}
        \toprule
        Work & Inference & FT & Model Sharing & Client Independence & Flexible Placement & Model Transparency & Batched Inf-FT & Multiple PEFT & Adapter Privacy \\
        \bottomrule
        vLLM~\cite{noauthor_vllm-projectvllm_2024} & $\checkmark$ & $\times$ & $\checkmark$ & $\times$ & $\times$ & $\checkmark$ & $\times$ & $\checkmark$ & $\times$ \\
        \hline
        mLoRA~\cite{mLoRA2023} & $\times$ & $\checkmark$ & $\checkmark$ & $\times$ & $\times$ & $\times$ & $\times$ & $\times$ & $\times$ \\
        \hline
        {\bf \scheme{}} & $\checkmark$ & $\checkmark$ & $\checkmark$ & $\checkmark$ & $\checkmark$ & $\checkmark$ & $\checkmark$ & $\checkmark$ & $\checkmark$ \\
        \hline
    \end{tabular}
    \caption{Design goals achieved in \scheme{} and other state-of-the-art multi-adapter systems. (FT: Fine-tuning)}
    \label{tab:sota}
\end{table*}

As shown in Table~\ref{tab:sota}, none of the existing multi-adapter systems achieve all aspects of the desired goals.

\begin{figure}[t]
    \centering
    \includegraphics[width=\linewidth]{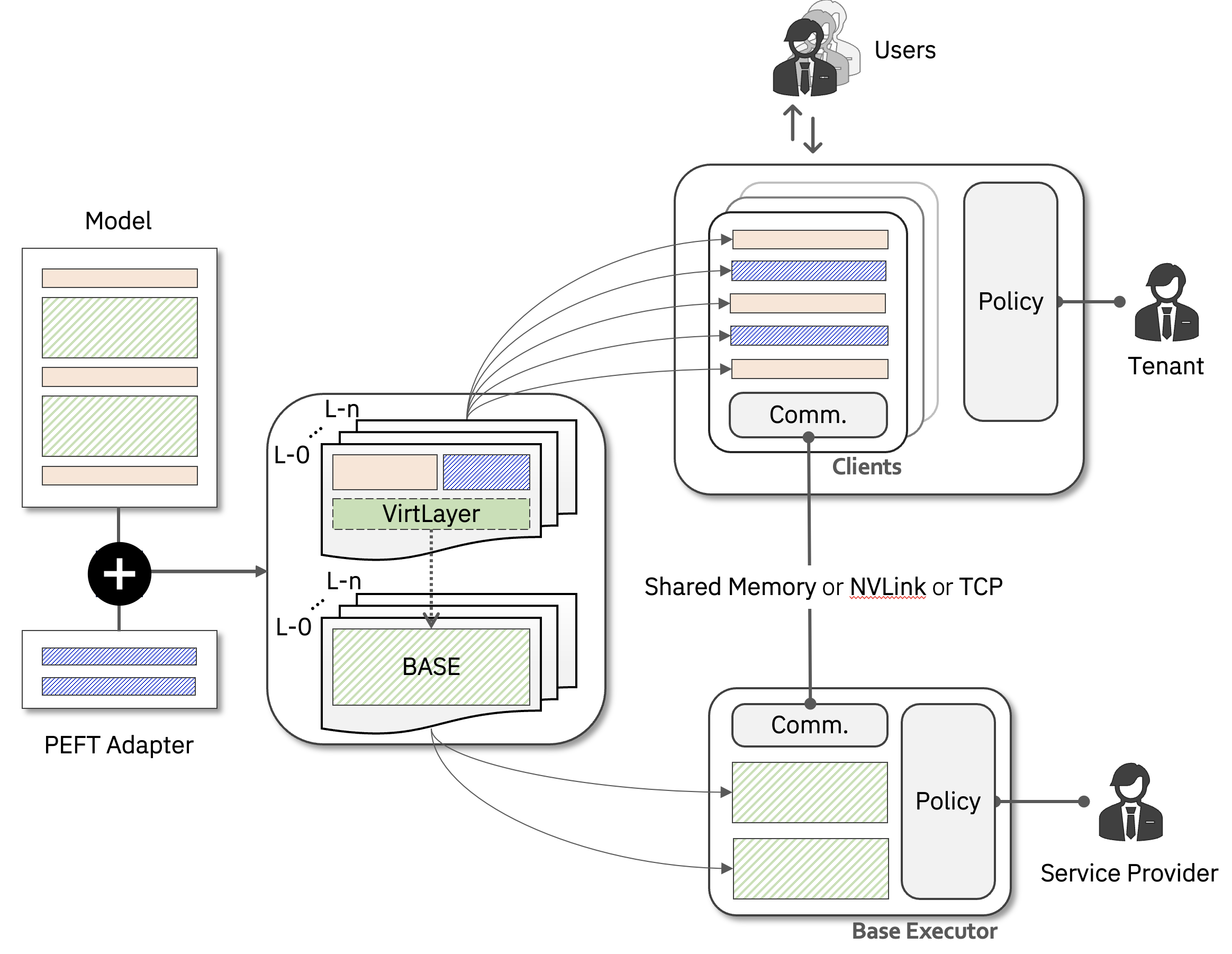}
    \caption{High-level design of \scheme{}}
    \label{fig:symbiosis-hld}
\end{figure}

\begin{figure}[htp]
    \centering
    \includegraphics[width=1.02\linewidth]{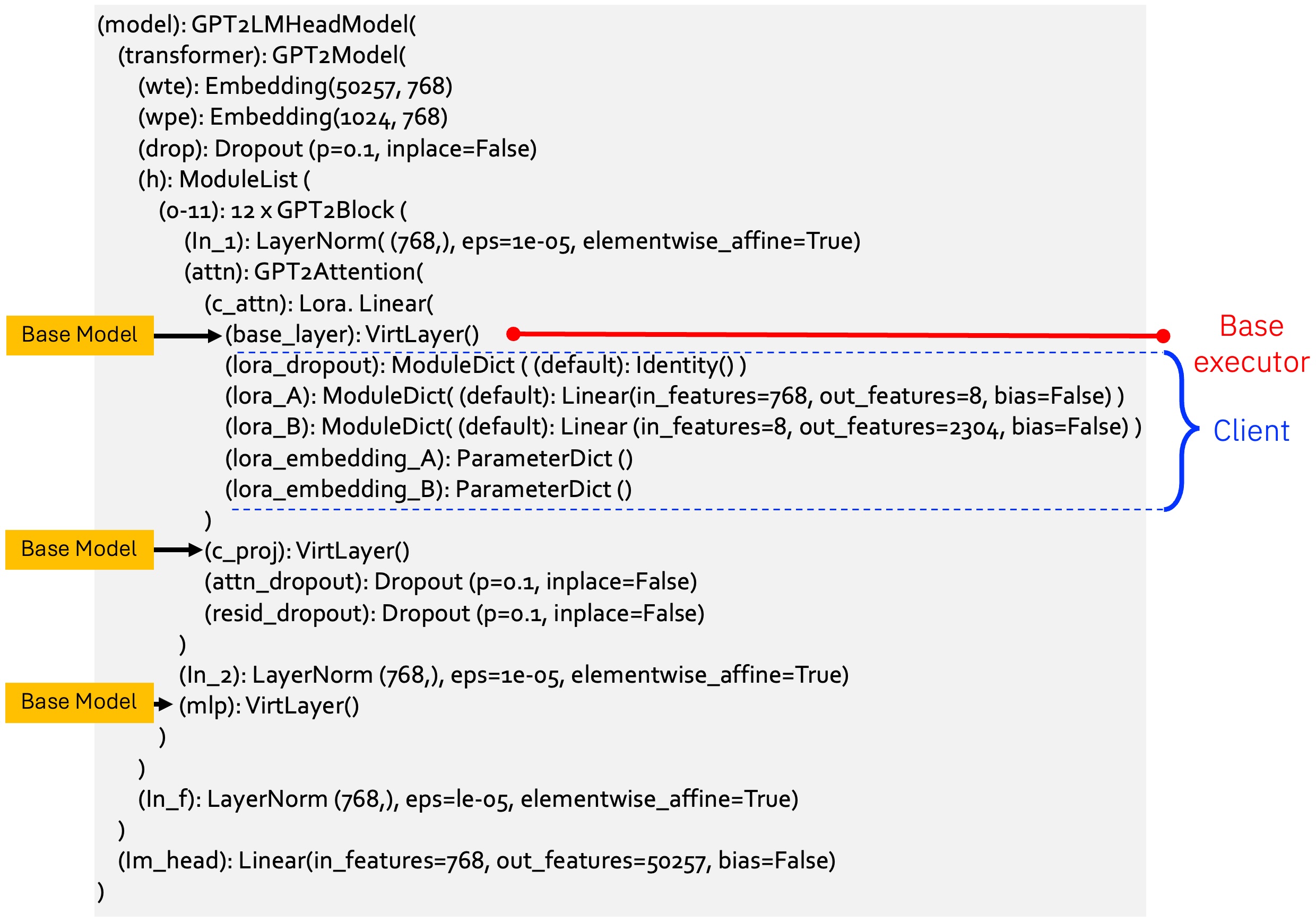}
    \caption{\scheme{} replaces the base model layers in the model structure on the client side with VirtLayer, which redirects their execution to the base executor.}
    \vspace{-1em}
    \label{fig:parse-model}
\end{figure}

\begin{figure*}[htp]
    \centering
    \includegraphics[width=0.8\linewidth]{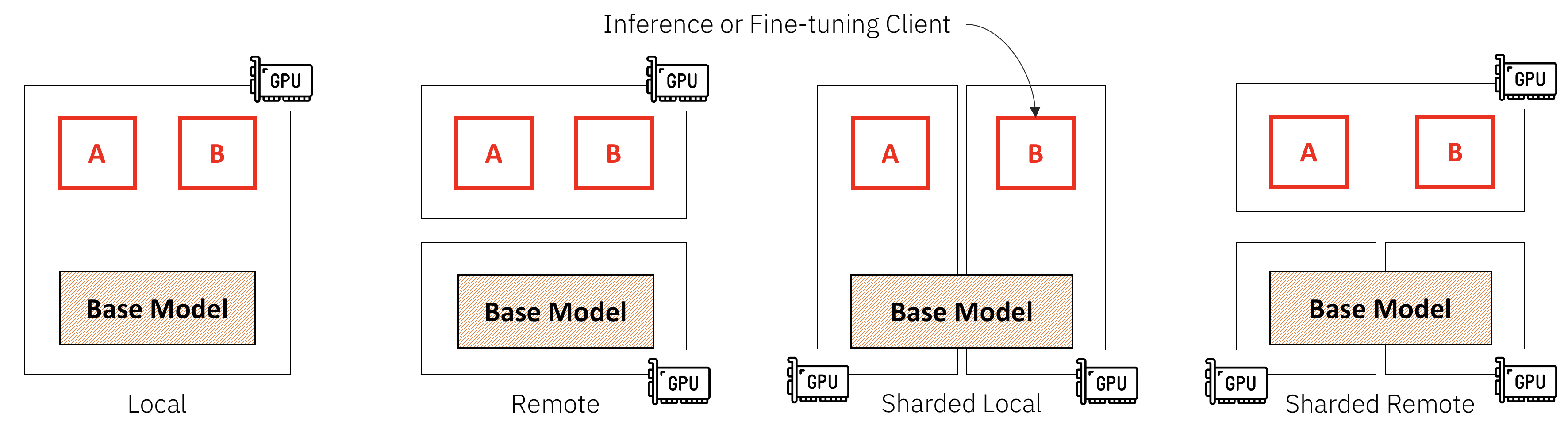}
    \caption{The figure shows the various possible configurations of base executor and clients in \scheme{}. Even though Remote and Sharded Remote configurations show clients on the same GPU, they can also be placed on different GPUs or nodes.}
    \vspace{-1em}
    \label{fig:sym-config}
\end{figure*}


\subsection{Split Execution in \scheme{}}
\label{sec:flow}
\paragraph{Design Overview}
Figure~\ref{fig:symbiosis-hld} shows the high-level architecture of \scheme{}. When loading the model, \scheme{} separates the base model layers as configured by the service provider, and loads them to the base executor. Whereas the adapter and non-base layers, such as attention and batch normalization layers, are loaded by each tenant into respective client. 
The client and base executor can communicate through various mechanisms shown in Figure~\ref{fig:symbiosis-hld} based on their host devices. The service provider defines the batching policy for inference and fine-tuning requests. Tenants can enforce their resource management policies on the GPUs (or nodes) they have been allocated with the help of off-the-shelf schedulers~\cite{llm-sched, llm-sched2, llm-sched3, llm-sched4}. This allows them the flexibility to define per-client resource constraints (e.g., maximum context length, batch sizes) and reflect their priorities for different types of jobs. Section~\ref{sec:privacy} discusses tenant-specific privacy configuration.

\paragraph{Client}
A client can be a trainer (in case of a fine-tuning job) or an inference client (in case of an inference job) and serves as an endpoint to receive the training data or requests. Each client selects a PEFT method, with its desired parameters, which \scheme{} should implement through the adapter layers. This creates an instance of client-specific non-base layers, while the base model layers are served by the common base executor. For an input from a user, a client processes all client-side non-base layers locally and invokes the base executor for the base model layers. Whenever the client encounters a base model layer in the model, it sends the corresponding activations to the base executor for processing. Upon receiving a response from the base executor, the client continues with the execution of the client-side layers until it encounters the next base model layer. Both the forward and backward passes follow this layer-wise execution. Eventually, the result is returned to the user (for inference), or an optimizer is invoked for model parameter update (for fine-tuning). Since each client controls the rate at which it sends activations to the base executor, clients can drive their inference or training independently. This allows \scheme{} to accommodate different rates of execution for different adapters. For instance, client A can perform twice as many iterations as client B, where client A may share its execution of the base model layers with client B for a fraction of iterations.





\paragraph{Base Executor}
An invocation of a base model layer on the client-side results in an invocation of the base executor. The base executor serves each base model layer separately, so they can be invoked independently by different clients. The client-side passes the activations for the base model layer to the executor as tensors. The base executor batches the requests for a layer received from multiple clients and executes the requested layer’s forward or backward pass. The output of this batch processing is then split into individual outputs and sent to the respective clients. In a forward pass of a fine-tuning job, the input and output tensors are saved to calculate the gradients in the corresponding backward pass. No tensors are saved for inference. \scheme{} introduces techniques (discussed in Section~\ref{sec:sched}) to reduce, and in some cases eliminate, the storage required to store the input and output tensors.

\paragraph{\scheme{} VirtLayer}
To redirect the execution of the base model layers to the base executor, \scheme{} scans and replaces the frozen layers (base model layers) in the client-side model definition with a custom {\em VirtLayer}. Like all the layers in PyTorch, VirtLayer is an instance of {\em torch.nn.Module} with custom built forward and backward functions. These functions have the same properties, return datatypes, and sizes, as the corresponding functions of the replaced base model layer. However, these functions don't execute the layers locally. Instead, when these functions are invoked, they send the necessary metadata, e.g., client id, target base layer, and the activation tensors to the base executor for processing. Upon receiving the response from the base executor, the functions return the received activations (forward pass) or gradients (backward pass) and the execution of the client-side layers continue. Each VirtLayer also has a base model layer identifier and other information needed to communicate with correct base model layer at the base executor. For example, VirtLayer contains client and base executor GPU identifiers (PyTorch ranks) in case of GPU-GPU communication. Since VirtLayer replaces the base model layer in the client-side model definition, this modification is performed through \scheme{} library call, eliminating the need to change the model code in the transformer library~\cite{transformers}. Figure~\ref{fig:parse-model} shows the modified model definition.

\subsection{Flexible Placement}
\label{sec:placement}
\scheme{}'s split execution decouples client and base executor resources, allowing for a flexible placement across different devices (or types of devices). Figure~\ref{fig:sym-config} shows the placement configurations supported by \scheme{}. The first configuration, local, show the case when one or more clients are located on the same GPU as the base executor. This configuration allows for fast communication while sharing the memory resources between clients and the base executor. In the remote configuration, clients are located on a different device than the base executor. The clients can be on another GPU on the same or a different node. It allows clients to scale independently of the base executor, where single or multiple GPUs can host several clients for the same base executor.

\scheme{} also allows sharding models across GPUs. Sharding splits layers across GPUs to reduce the memory footprint per GPU. Whenever a layer is executed in base executor, only the parameters corresponding to that layer are fetched from all the GPUs. After the layer's execution, the fetched parameters are released, freeing the memory. The sharded local configuration allows scaling of the base model across multiple GPUs by sharding the base layer weights across them. A client can be present in any one of the GPUs where a shard of the base model resides. The base executor provides a communication endpoint at each of the GPU where its layers are sharded. Hence, a client only needs to communicate with its local shard. Lastly, the sharded remote configuration is similar to the remote configuration except that the base model layers are sharded across GPUs. This configuration allows executing the largest models, where both base executor and the clients can scale independently.

To realize sharding, we utilize Fully-Sharded Data Parallelism (FSDP) to shard the base model layers. FSDP enables sharding of parameters, gradient synchronization, and data-parallel training. Since all our base model layers are frozen (not trainable), we only use the sharding capabilities of FSDP. We design a custom FSDP wrapping strategy that marks individual base layers as independent FSDP instances. This enables base layers to scale and execute independent of the client layers, hence decoupling the execution of different layers.


\subsection{Long-Context Inference with Heterogeneous Compute}
\label{sec:hg-inf}
\scheme{}'s flexible placement and heterogeneous compute capabilities can be used to design an efficient system for long context inference.
A significant fraction of inference runtime state in decoder models is the key-value (KV) cache, containing the key and value states used for calculating attention. The size of this KV cache can be large (for example, 8GB for Llama2-7B model with an input sequence length of 16K and a batch size of just 1). Moreover, it grows linearly with input sequence length and batch size, often exceeding the memory available on a GPU.
The state-of-the-art solutions offload the KV cache to CPU (exploiting the comparatively larger host memory), and saving GPU memory at the expense of increased CPU-GPU transfers~\cite{transformers}. However, for large output sequence lengths and/or batch sizes, the overhead of CPU-GPU transfers can become worse than the benefits of GPU acceleration as shown later in Figure~\ref{fig:cpu-gpu-hg-inf}. 

\begin{figure}[t]
    \hspace*{-1em}
    \includegraphics[width=1.07\linewidth]{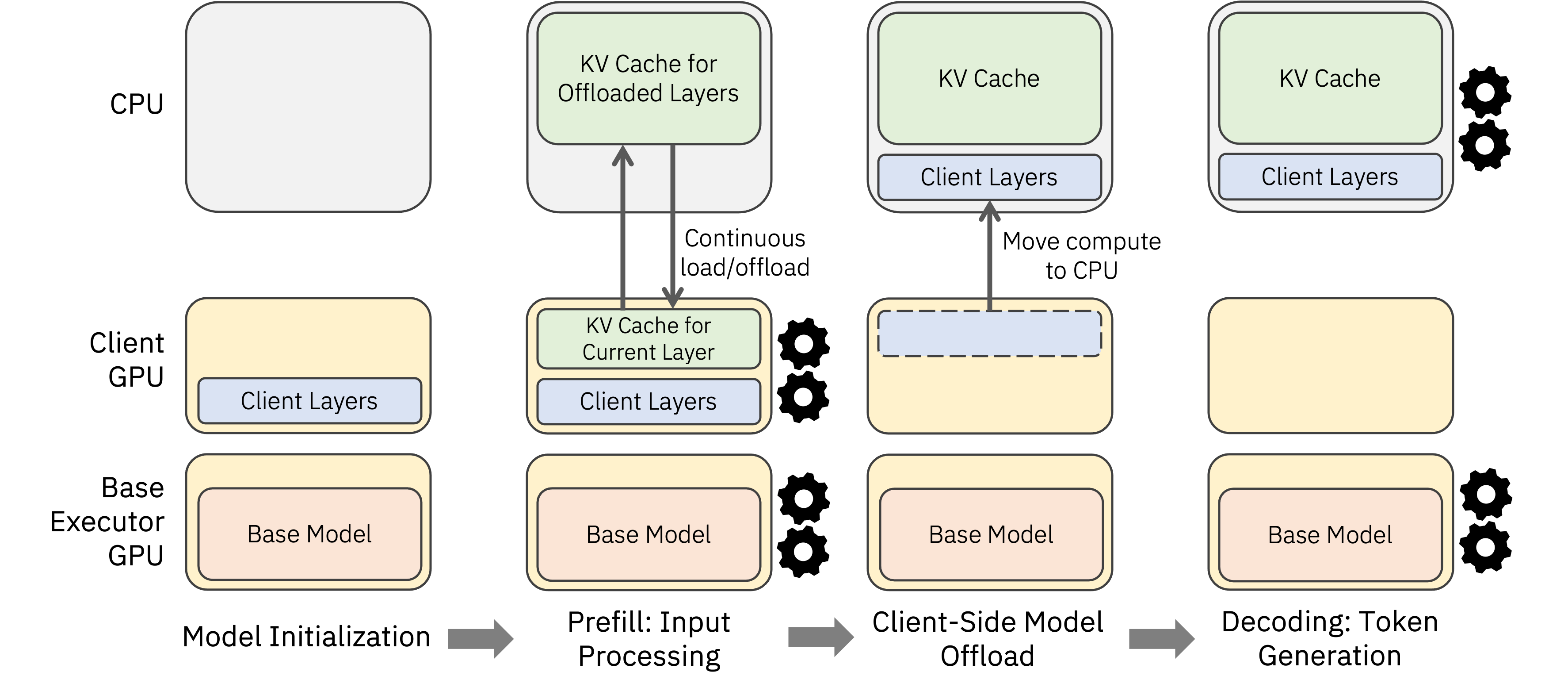}
    \caption{Inference for large inputs in \scheme{} using flexible placement and heterogeneous compute. The wheels mark the devices performing computations during a stage.}
    \label{fig:hg-inf}
\end{figure}

Figure~\ref{fig:hg-inf} shows the execution of inference in \scheme{}. While we show client and base-executor are located on different GPUs, any of the placement configurations discussed earlier can be selected. After the client layers and base model are loaded into the respective devices, we perform the throughput-sensitive prefill on GPU. The prefill stage processes the entire input prompt, therefore it requires GPU acceleration to complete the prefill reasonably quickly. During prefill, we use the OffloadedCache feature proposed in ~\cite{offloadedcache} to offload the KV cache to CPU memory. 
Prefill is followed by the decoding stage, which generates one output token at a time. Since the KV cache is already offloaded to the CPU memory, the client-side layers are also loaded and executed on the CPU. We show in Section~\ref{eval:cpu-gpu-hg-inf} that such heterogeneous compute is faster than all-GPU compute, where KV cache needs to be transferred from CPU memory to GPU memory for generation. Moreover, since GPU cannot accommodate the entire KV cache, every iteration, the executing layer's KV cache is fetched right before their execution.

\subsection{Client - Base Executor Communication}
\scheme{}'s communication mechanics depends upon the relative placements of the client and base executor.
When the client and base executor are located on different GPUs, the communication occurs over {\em nccl} protocol, where tensor is transferred between the client and the base executor GPUs. {\em nccl} can utilize GPU kernels to allow fast intra-node GPU-GPU communication over NVLink for supported devices. However, nccl does not support exchanging tensor between client and base executor if they are co-located on the same device.

When client and base executor are located on the same GPU, \scheme{} implements a local communication mechanism. The clients and base executor communicate metadata and control messages over TCP via ZeroMQ~\cite{zeromq}, whereas the data is transferred through a shared tensor. Sharing obviates the need to transfer or copy the data and reduces communication latency. Sharing tensor across the processes is performed by acquiring the tensor metadata using {\em share\_memory\_()} method in the source process and rebuilding the reference using {\em rebuild\_cuda\_tensor()} method in the target process. However, the expensive CUDA calls for each layer can substantially increase the communication latency. To avoid this communication penalty, we pre-allocate a shared tensor for each client. This pre-allocated tensor is used for exchanging all input/output tensors between a server and a client.

Upon initialization, each client allocates a tensor of size (batch size) $\times$ (sequence length) $\times$ max(input, output dimension) using pre-determined values for all dimensions. Using the maximum of input and output dimension of a model as the last dimension allows the same tensor to be used during forward and backward pass of any layer. If the tensor batch size or sequence length is insufficient for the requests, the shared tensor is resized to accommodate the desired batch size and sequence length.




\begin{figure}[htp]
    \centering
    \includegraphics[width=0.9\linewidth]{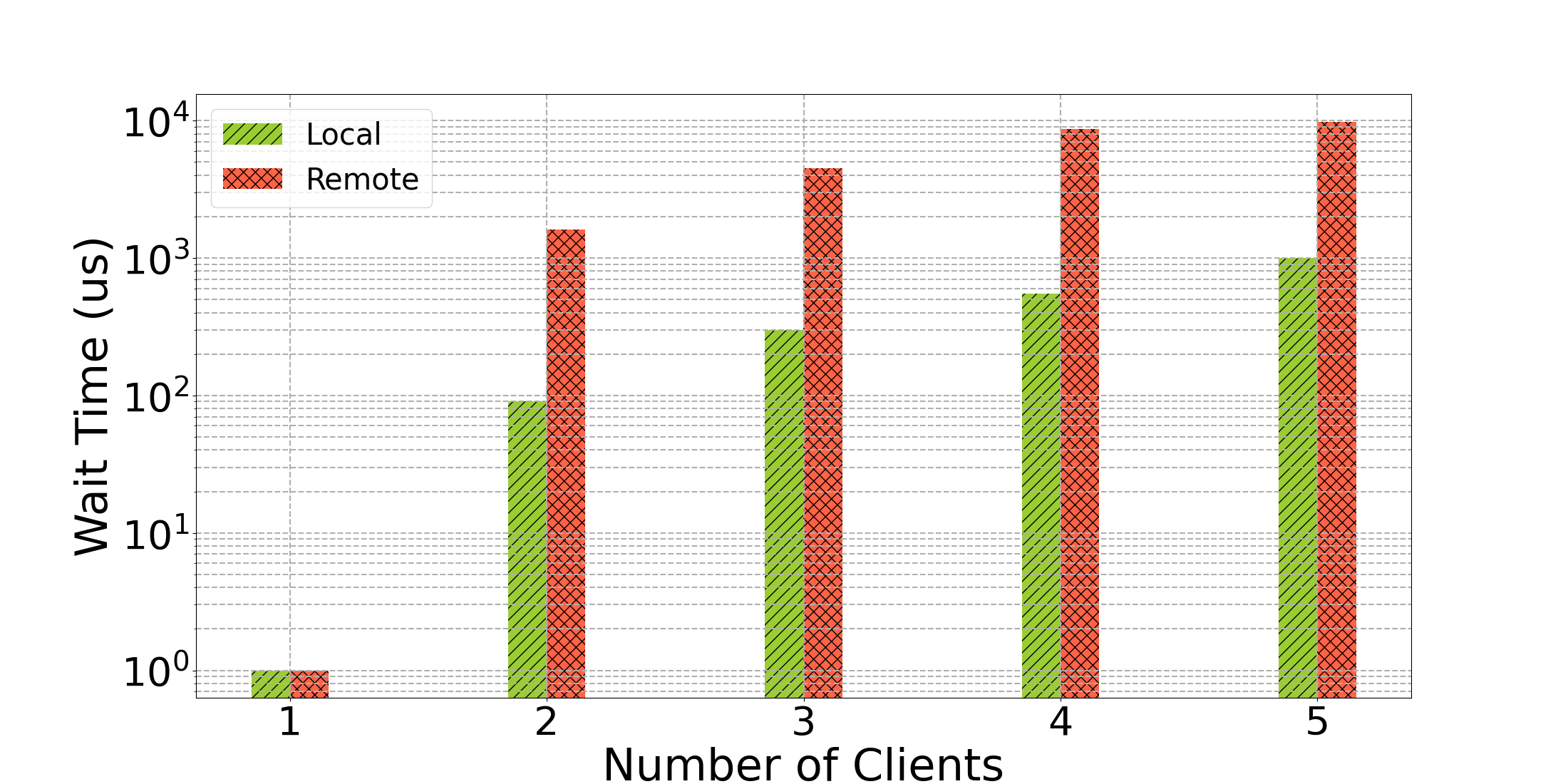}
    \caption{Per-layer wait time at the base executor for local and remote inference clients of Llama2-7B.}
    \label{fig:inf-wait-lockstep}
\end{figure}

\subsection{Independent Execution}
\label{sec:sched}
PyTorch requires that the requests that were batched together for the forward pass of a layer be also be batched during their backward pass. This means that if inference and fine-tuning requests were batched together, both requests need to perform a backward pass. The inference requests do not require the backward pass and unnecessarily performing a backward pass would waste the GPU compute cycles. The limitation also means that if any fine-tuning requests that were batched together for a forward pass of a certain layer, should also be batched together for the backward pass of that layer. This limitation emerges from PyTorch's implementation. When performing a forward pass, PyTorch captures the identity of the input/output tensors and the corresponding operation in a computation graph. This computation graph is traversed during the backward pass to compute gradients. PyTorch requires that the same tensors used in the forward pass to be used in the backward pass for gradient calculation. Therefore, the popular platforms, such as Transformers ~\cite{transformers}, address this problem through {\em lockstep execution}, where a batch is maintained through the execution of forward and backward passes for all the layers.

However, this prevents independent execution of different requests. With lock step execution, the execution of each layer at the base executor needs to wait for all the previously batched clients to complete their client-side execution, e.g., attention. However, when using \scheme{} to serve base model as-a-service, the service provider may receive a variety of requests, where each client requires different time for the client-side computation. This can be because of the differences in client configuration (e.g, LoRA rank), hardware (e.g., GPU or CPU), or location (e.g., local vs. remote). Batching such diverse requests for all layers results in performance degradation. Figure~\ref{fig:inf-wait-lockstep} shows the per-layer wait time for local and remote client configurations of \scheme{} with lockstep execution.

In \scheme, we eliminate the above batching requirement using the following two insights. First, since the layers on the base executor are frozen the parameters are not updated during a backward pass. Second, for the most popular LLM architectures, (e.g., Llama, GPT, Gemma, Bert), linear and 1D convolution layers constitute the largest fraction of the model parameters. For these layers, the input and output tensors are not involved in gradient calculations. Specifically, the gradient of output w.r.t. input resolves to the parameters themselves. Hence, \scheme{} does not store the input/output tensors for linear and 1D convolution base model layers, even for the fine-tuning request. Instead, it performs a matrix-multiplication between the output gradients and the parameters to generate the required gradients for these layers during the backward pass. This breaks the lockstep, while also providing significant memory savings by not requiring to store input/output tensors for each client at the base executor.

\subsection{Opportunistic Batching}
While breaking the lockstep execution frees each client to execute independently, this leads to smaller batches and requires the base executor to perform more iterations. To address this challenge, \scheme{} allows the base executor to wait to exploit the opportunity to accumulate new incoming requests and create larger batches. This is referred to as {\em opportunistic batching}.
To honor the latency of latency-sensitivity of inference request, we allow them to progress faster through the model without having to wait for less latency-sensitive requests for better batching. To accomplish this, we base the wait time on the size of request.
For instance, fine-tuning, prefill or a large batch of inference requests can afford to wait longer than smaller requests, since the wait time is a smaller fraction of their naturally longer iteration latency.

\scheme{} also leverages the insight that for $Conv.1D$ and $nn.Linear$ layer computation, the position of token does not matter in a sequence. Therefore, we are able to flatten all {\em batch-size X sequence-length} inputs received from different clients into 1-dimensional sequence of tokens. This allows \scheme{} to avoid padding (which is required to accommodate different size inputs) and save wasted computation.

\subsection{Privacy for Multi-Tenancy}
\label{sec:privacy}
\begin{figure}[h]
    \centering
    \includegraphics[width=0.6\linewidth]{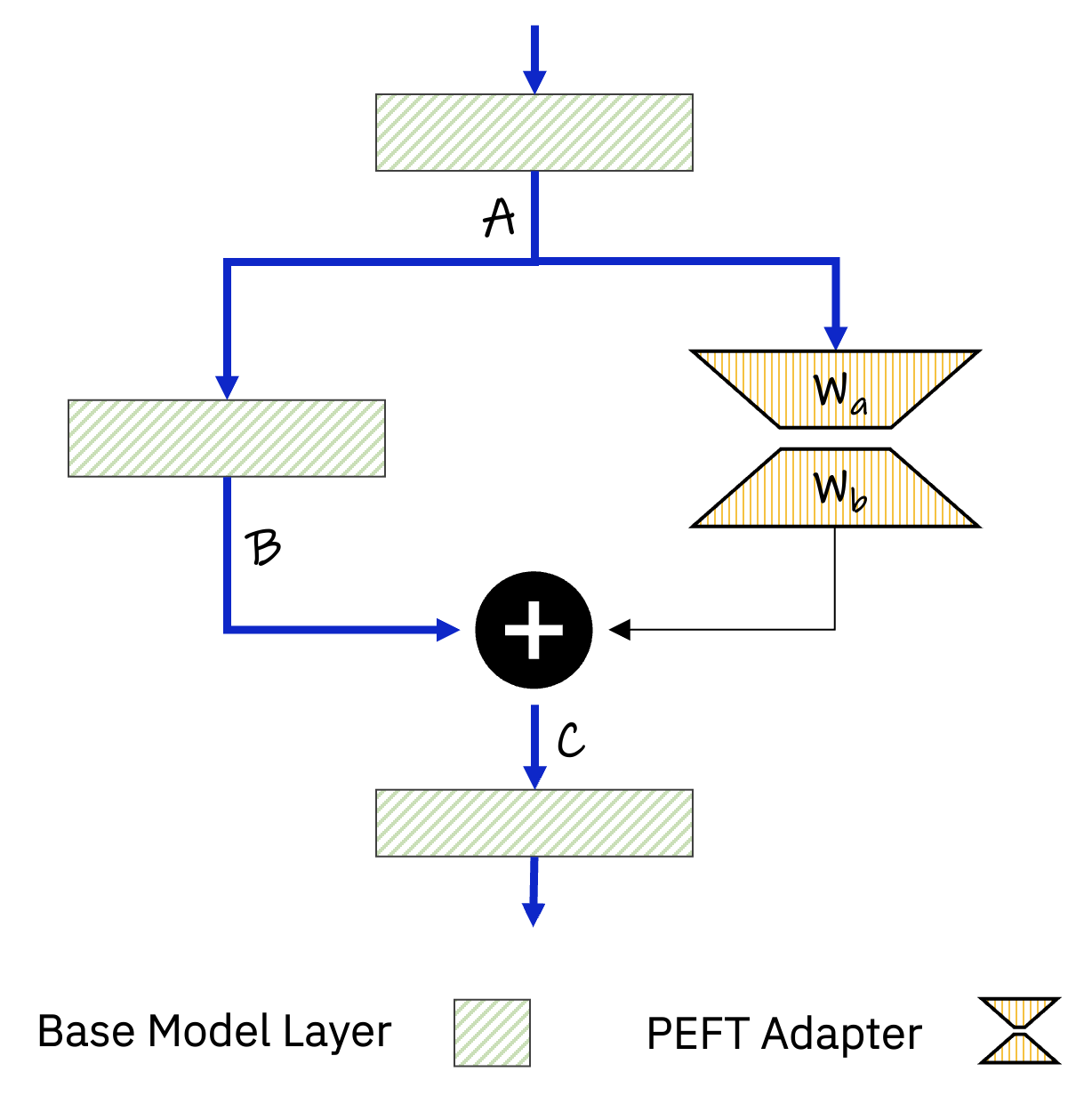}
    \caption{In \scheme{}, the base executor has access to the activations marked in blue colored lines. Therefore, the effect of adapter parameters ($Wa$ $.$ $Wb$) can be computed with $(C - B) / A$.}
    \label{fig:privacy-lora}
\end{figure}
In a multi-tenant environment, we consider the threat model where the adversary's goal is to extract the adapter parameters. The tenants that have trained their adapter on confidential data may not wish to expose their adapter parameters and request-specific run-time state (KV cache) to the infrastructure provider or the base executor service provider. In \scheme{}, the third-party base executor can observe the client-side activations and steal its functionality, like the model extraction attacks~\cite{honeypotnet, me-attack, me-attack2, me-attack3}. Figure~\ref{fig:privacy-lora} shows an example of the model architecture where the base executor can extract LoRA adapter parameters in \scheme{}.

We address the privacy concern in the following ways. First, \scheme{} decouples the tenant specific state into a client process. Therefore, the client process can be hosted on a secure host deployed in tenant's environment and it can communicate with the base model service over the network. Second, to address the privacy of activations, client adds noise to the activations before sending them to the base executor and the effect of noise is subtracted from the base executor's output. {\em Note that due to subtraction, this mechanism does not alter the overall result, i.e., the model produces the exact output which it otherwise would have in a non-privacy preserving setting.} To accomplish this, we first send the noise to the base executor to calculate the {\em noise effect}. During regular execution, the base executor produces {\em noisy output}. The noise effect is subtracted from this noisy output to generate the actual output. The noise effect need not be calculated for every iteration, but only once for a given noise value. To further prevent the leakage of noise, the tenant can either periodically change the noise or prepare several noise values in advance and pick different values for different iterations.

The calculate the effect of noise, we leverage the insight that common LLMs base layers are either $nn.Linear$ or $nn.Conv1D$ layers. For the layers that do not follow linearity, such as $Sigmoid$, $Softmax$, it is not possible to separate the effect of noise from the noisy output and produce the original output. Another challenge with this privacy preserving technique is the presence of bias in $nn.Linear$ or $nn.Conv1D$ layers, which prevents the calculation of equation 2 below. To address this, we also host an alternate execution flow in the base executor which nullifies the effect of bias and returns $n_{effect}$.

\begin{enumerate}
    \item Calculation of noise effect.
\[n_{effect} = Conv1D(n, W)\]

    \item Calculation of noisy output by adding $n$ to input $x$ at client before forwarding it to the base executor.
\[y_{noisy} = Conv1D(x + n, W) + b\]
For Conv1D or nn.Linear, the above can be expanded to,
\[y_{noisy} = Conv1D(x, W) + Conv1D(n, W) + b\]

    \item Calculation of actual output by removing the effect of noise.
\[y = y_{noisy} - n_{effect}\]
\end{enumerate}

\begin{itemize}
    \item W: Parameters
    \item b: Bias
    \item x: Original input
    \item y: Desired output
    \item n: Noise
    \item $y_{noisy}$: Noisy output
    \item $n_{effect}$: Effect of noise on output
\end{itemize}

Since the base executor has access to the noise that is forwarded for calculating the $n_{effect}$, different unique noise values can be used for different layers. This information is only known to the tenant. With only 2 noise values, correctly guessing the noise value used by the client for all layers (e.g., Llama2-7b contains of 100s of nn.Linear layers) is difficult because of extremely large number of possible combinations. Also note that the privacy preservation itself does not address the security of the client environment, where users can rely on existing methods, such as NVIDIA MIG~\cite{mig}, to create a secure environment.


\section{Evaluation}
Our evaluation test-bed consists of 8 NVIDIA A100 GPUs each with 80GB of memory. The host has 64 AMD EPYC 7763 CPU cores and 512GB of memory. Table~\ref{tab:models} list the models used in the following experiments. We compare \scheme{} with the following popular inference and fine-tuning platforms. Note that \scheme{} is not an inference or fine-tuning platform by itself, it derives its performance from the optimizations in the underlying Transformers library.
\begin{itemize}
    \item {\em Baseline:}  Method provided by Transformers~\cite{tx} to fine-tune a single adapter. For fine-tuning an adapter using multiple GPUs, we use FSDP as the baseline.
    \item {\em mLoRA:} An open source tool for simultaneous fine-tuning of multiple LoRA adapters.
    \item {\em vLLM:} An inference platform, which allows the base model to be shared across multiple PEFT adapter.
\end{itemize}

For workload, we generate randomly initialized input tensors of desired batch size and sequence lengths. Since, the output with \scheme{} is exactly identical to that of the baseline, the content of the input is not relevant to the performance metrics below.
For most fine-tuning experiments, we use a batch size value of 2, a common default configuration across popular platforms~\cite{aws-bs, ibm-bs, unsloth-bs}. This is because the GPU memory consumption is proportional to the product of number of clients, batch size and sequence length. Therefore, smaller batch sizes are best suited to demonstrate the performance with longer sequences and increasing number of clients.

We use LoRA adapters for the following experiments. However, \scheme{} supports other fine-tuning methods such as IA3 and prefix tuning. Table~\ref{tab:lora} shows the performance of fine-tuning a LoRA adapter with different configurations (Rank, Fine-tuned layers). As compared to the LoRA rank, addition of fine-tuned layers contributes more to increased latency. Therefore, in the evaluation below, we use the LoRA3 adapter, which uses maximum possible LoRA layers, namely, q, k, v, o.

\begin{table}
    \centering
    \begin{tabular}{p{2.5cm}p{2.5cm}p{2.5cm}}
        \toprule
        Adapter & Baseline & \scheme{}\\
        \bottomrule
        LoRA 1 & 0.32 & 0.4\\
        \hline
        LoRA 2 & 0.33 & 0.46\\
        \hline
        LoRA 3 & 0.37 & 0.57\\
        \hline
        LoRA 4 & 0.4 & 0.68\\
        \hline
    \end{tabular}
    \vspace{1em}
    \caption{\footnotesize Fine-tuning iteration latency of LoRA adapters with Llama2-13B. {\bf LoRA 1}: (8, [q]), {\bf LoRA 2}: (64, [q]), {\bf LoRA 3}: (8, [q, k, v, o]), {\bf LoRA 4}: (64, [q, k, v, o])}
    \vspace{-1em}
    \label{tab:lora}
\end{table}

\begin{table}
    \centering
    \begin{tabular}{p{2.5cm}p{2.5cm}p{2.5cm}}
        \toprule
        Model & Size (GBs) & Number of Layers\\
        \bottomrule
        GPT2-XL & 6 & 48\\
        \hline
        Llama3-1B & 2 & 32\\
        \hline
        Llama2-7B & 13 & 32\\
        \hline
        Llama2-13B & 26 & 40\\
        \hline
        Granite-20B & 40 & 52\\
        \hline
        Starcoder-15B & 60 & 40\\
        \hline
        Gemma2-27B & 56 & 46\\
        \hline
    \end{tabular}
    \vspace{1em}
    \caption{\footnotesize Models used in the experiments below. This demonstrates the generality of \scheme{} with Llama, BigCodeGPT, Gemma, GPT, GPTBigCode architectures.}
    \vspace{-1em}
    \label{tab:models}
\end{table}

\subsection{Memory Consumption}
In this section, we compare the memory consumption of the baseline with \scheme{} for a single and increasing number of fine-tuning jobs on a single 80GB GPU.
\subsubsection{Single Fine-Tuning Job}
In Figure~\ref{fig:single-trainer}, we compare the memory consumption of \scheme{} with the baseline for fine-tuning of a single rank-8 LoRA adapter. The \scheme{} without memory optimized backward pass increases the memory requirement compared to baseline. This increase is from having to maintain two copies of input and output tensors (on client side and base executor side) for backward pass. Even though, we don't use the client side tensors for gradient computation, the client's computation graph keeps track of the tensors to perform the backward pass. 
\scheme{}'s Memory Optimized (MO) version addresses this problem by eliminating the need of the base executor-side tensor copy, thus making \scheme{}'s memory footprint similar to that of the baseline. Moreover, unlike the baseline, the optimization provides constant base executor-side memory footprint even with increasing sequence length.

\subsubsection{Multiple Fine-Tuning Jobs}
\begin{figure*}[h]
    \begin{minipage}{0.33\linewidth}
    \centering
    \vspace{0.5em}
    \includegraphics[width=0.8\linewidth]{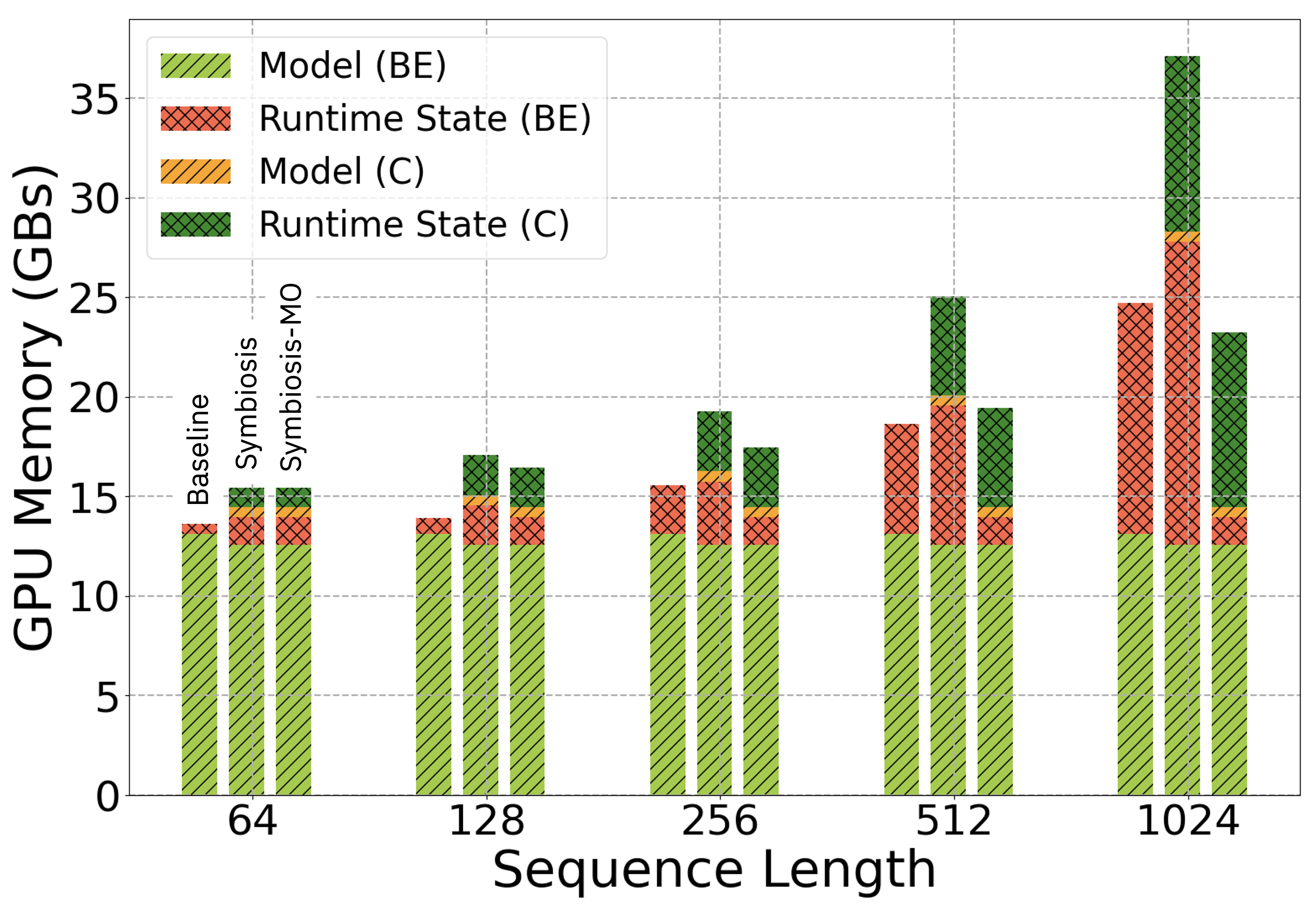}
    \caption{\footnotesize GPU memory consumption of \scheme{} for a file-tuning of a single rank-8 LoRA adapter. \scheme{}-MO represents memory optimized version.}
    \label{fig:single-trainer}
    \end{minipage}\hfill
    \begin{minipage}{0.32\linewidth}
    \centering
    \includegraphics[width=\linewidth]{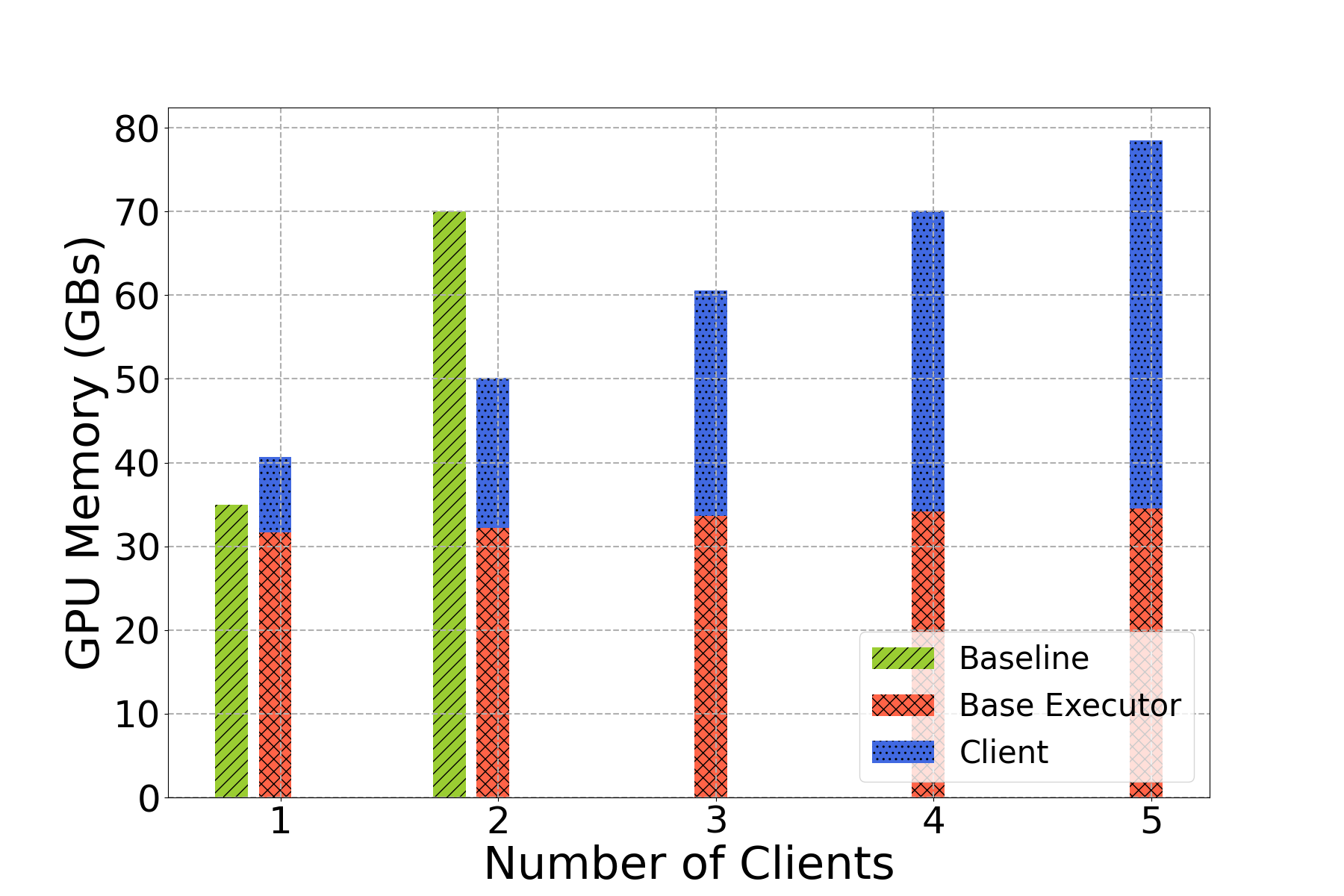}
    \caption{\footnotesize GPU memory consumption for Llama2-13B, batch size=2 and sequence length=512. The base executor memory footprint remains constant with increasing clients.}
    \label{fig:multi-trainer}
    \end{minipage}\hfill
    \begin{minipage}{0.32\linewidth}
    \centering
    \includegraphics[width=\linewidth]{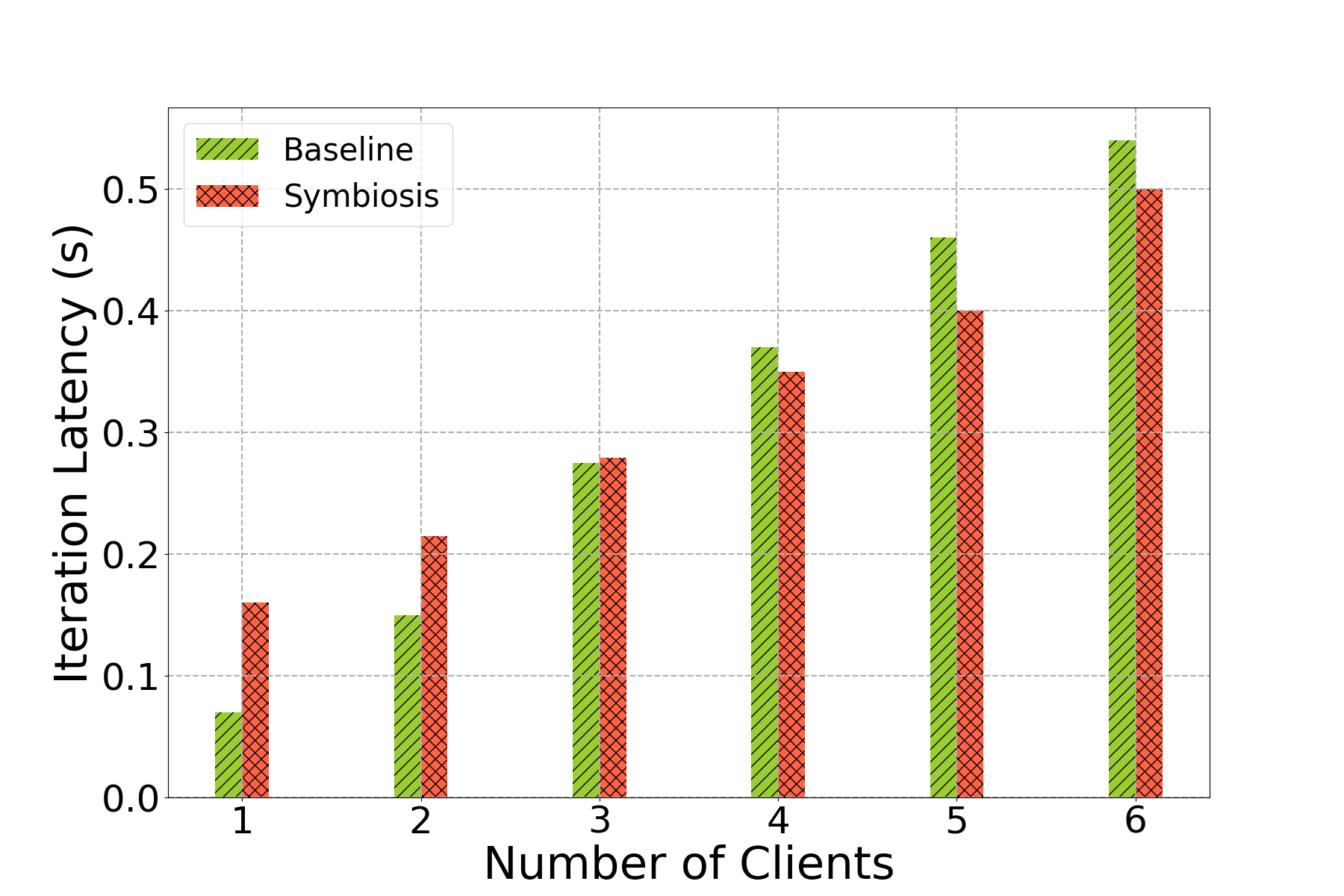}
    \caption{\footnotesize {\bf Single GPU:} Iteration latency for Llama3-1B, batch size=2 and sequence length=512. GPU contention hampers baseline latency more than that of \scheme{}.}
    \label{fig:ft-single}
    \end{minipage}\hfill

    \caption*{}
\end{figure*}

Figure~\ref{fig:multi-trainer} shows the memory consumption of GPU with multiple fine-tuning jobs. Since the input and output tensors are shared across multiple clients, we only observe a slight increase in base executor side memory consumption from having to use larger tensors. The client side consumption increases linearly with increasing clients as expected. From efficient use of memory, \scheme{} can accommodate 5 clients and the base model on a single GPU, whereas the baseline can only accommodate 2 independent fine-tuning jobs.

\subsection{Multi-Adapter Fine Tuning}
In this section, we present the evaluation of multi-adapter fine-tuning across single and multiple GPUs.
\subsubsection{Single GPU Fine Tuning}
Here we compare the performance of the baseline fine-tuning with \scheme{} for increasing LoRA fine-tuning jobs on a single 80GB GPU. For comparison with the baseline, we pick a smaller model, namely, Llama3-1B. From Figure~\ref{fig:ft-single}, it can be observed that the baseline outperforms \scheme{} in terms of latency up to 2 clients. However, beyond 2 clients, the lack of batching in baseline results in resource contention. In contrast, \scheme{} can better amortize the impact of client-base executor communication with cross-client batching and achieve lower latency. Figure~\ref{fig:ft-single-thr} shows the corresponding token throughput. Due to saturation of GPU's computational capability with increased batching, the throughput with \scheme{} starts to diminish at 6 fine-tuning clients. Note that since we use a smaller model for comparison with the baseline, the impact of communication is magnified as a fraction of the total iteration latency.

\subsubsection{Multi GPU Fine Tuning}
Here, we evaluate the fine-tuning performance of \scheme{} over multiple GPUs. For the sharded mode, we compare our fine-tuning performance with that of baseline and mLoRA. We use Llama for comparison because it is the only model supported by mLoRA.

\begin{figure*}[h]
    \begin{minipage}{0.32\linewidth}
    \vspace{-2em}
    \centering
    \includegraphics[width=\linewidth]{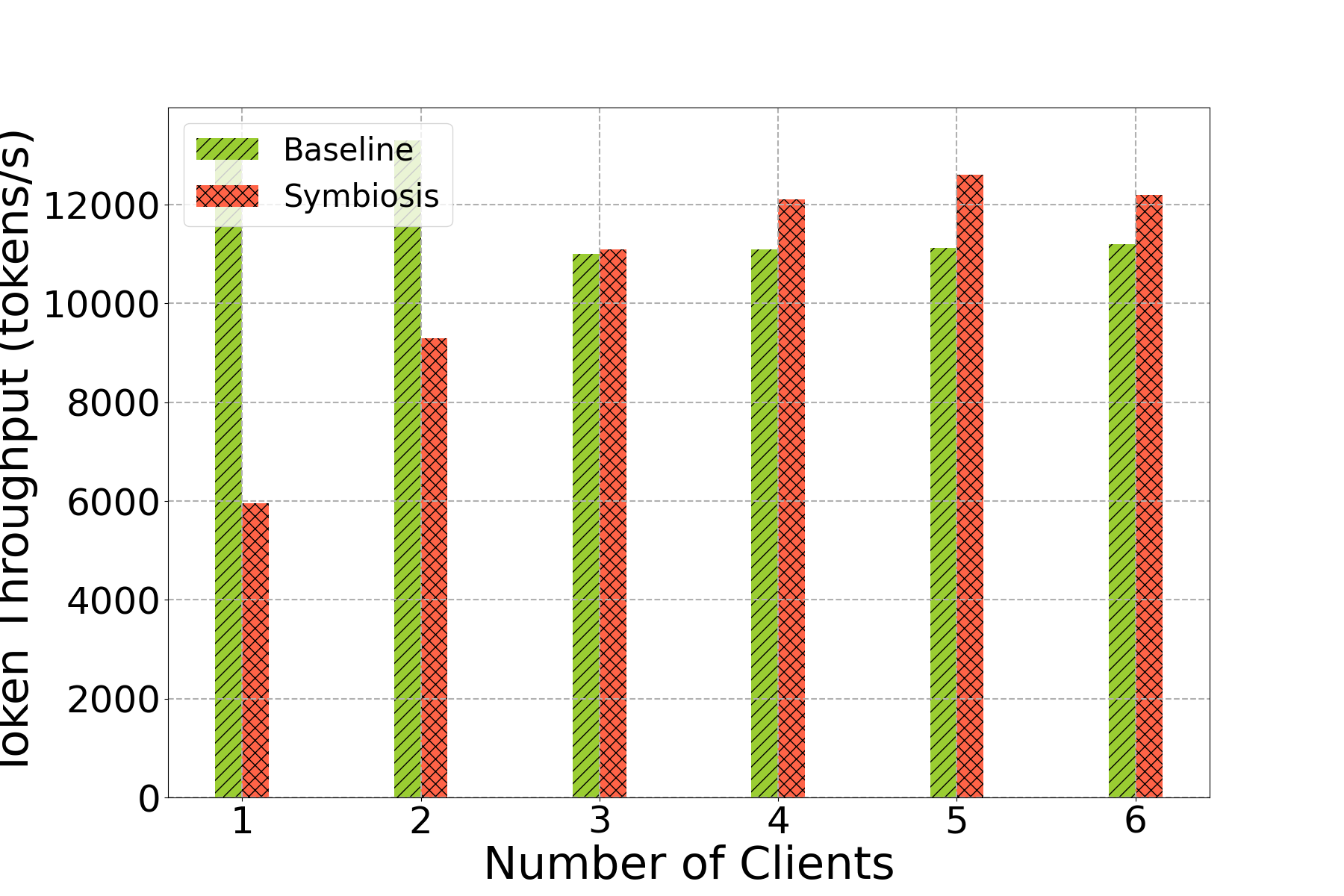}
    \caption{\footnotesize {\bf Single GPU:} Token throughput for Llama3-1B, batch size=2 and sequence length=512.}
    \label{fig:ft-single-thr}
    \end{minipage}\hfill
    \begin{minipage}{0.32\linewidth}
    \vspace{-2em}
    \centering
    \includegraphics[width=\linewidth]{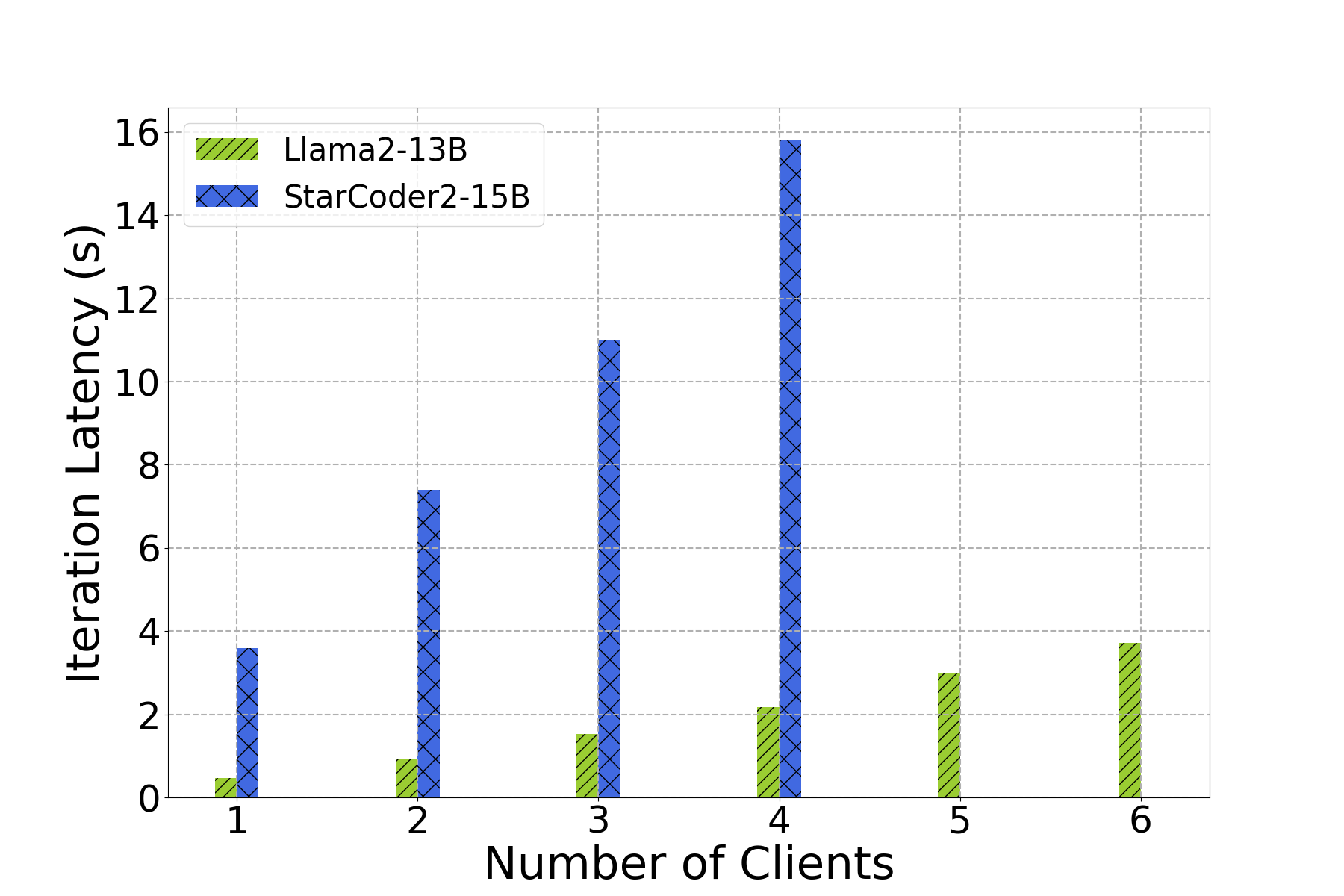}
    \caption{\footnotesize {\bf Remote Execution:} Iteration latency with batch size=2 and sequence length=512. 1 client GPU and 1 base executor GPU.}
    \label{fig:ft-remote-2-gpu-lat}
    \end{minipage}\hfill
    \begin{minipage}{0.32\linewidth}
    \vspace{-2em}
    \centering
    \includegraphics[width=\linewidth]{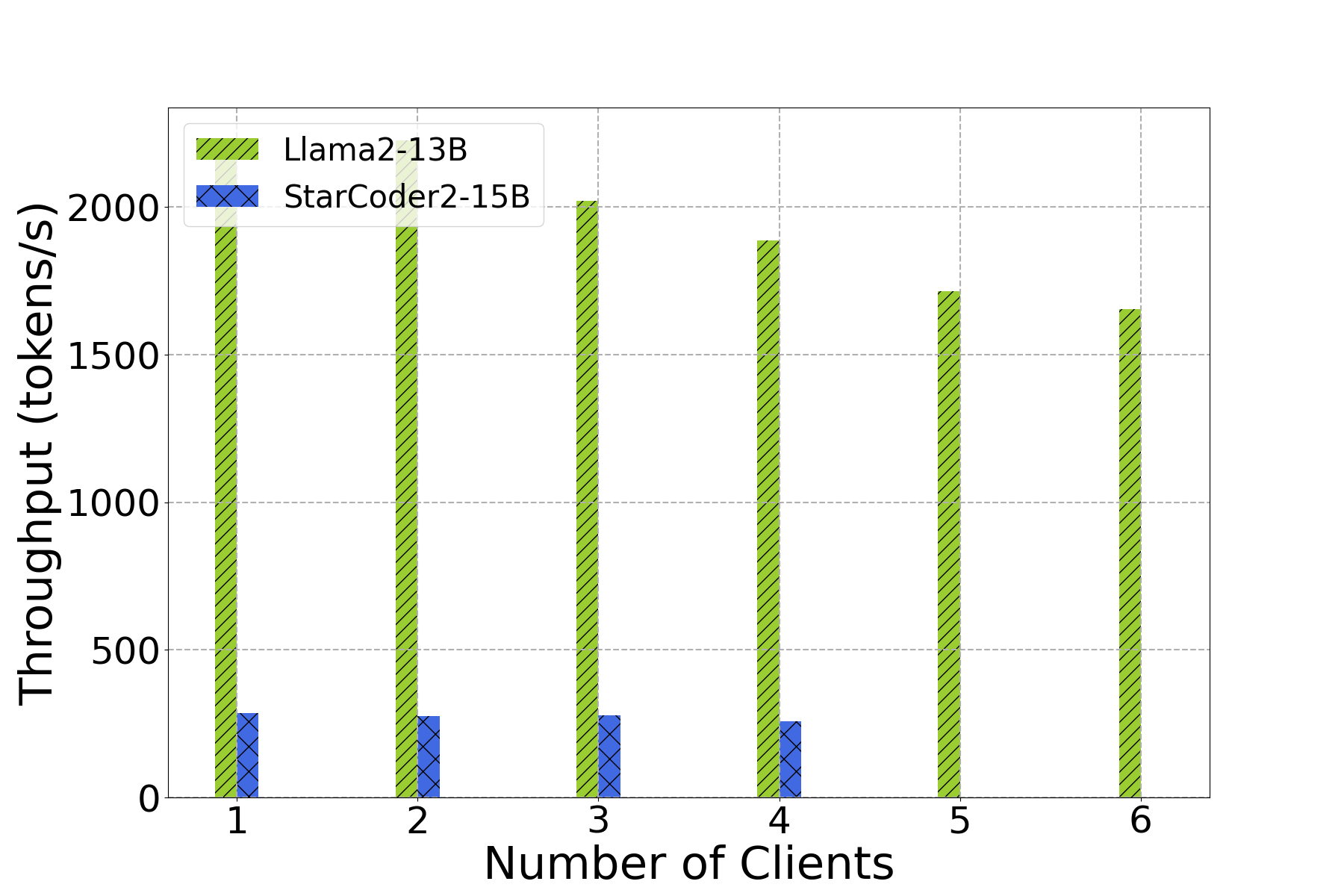}
    \caption{\footnotesize {\bf Remote Execution:} Token throughput with batch size=2 and sequence length=512, 1 client GPU and 1 base executor GPU.}
    \label{fig:ft-remote-2-gpu-thr}
    \end{minipage}\hfill
    \caption*{}
\end{figure*}

\paragraph{Remote Execution}
For remote execution, we host the base executor on one GPU and run clients run on another GPU. The clients and base executor communicate through NVLink. This configuration separates the base executor from clients, allowing clients to spread across several GPUs. Such configuration is best for fine-tuning with large sequence lengths, where client memory footprint from the runtime state is significant. Moreover, this configuration protects the server from clients with variable resource consumption.


For this 2 GPU experiment, we run all clients on a single GPU and the base executor on another GPU. Figures~\ref{fig:ft-remote-2-gpu-lat} and ~\ref{fig:ft-remote-2-gpu-thr} show per-iteration latency and throughput with increasing number of fine-tuning clients.  For \scheme{}, since the clients communicate across the GPUs, our latency and throughput is worse than local configuration. For Llama2-13B, we also observe increasing communication overhead with increasing clients. The fine-tuning performance of Starcoder2-15B is much worse than Llama2-13B from its larger size (60GB). Also, its 32-bit precision requires order of magnitude longer time for common operations, such as matrix multiplication~\cite{ft-precision}, compared to 16-bit precision. The single GPU (baseline) with Starcoder2-15B also requires 3.3s for a fine-tuning iteration with 310 tokens/s throughput (batch size=2 and sequence length=512). 



\begin{figure*}[h]
    \begin{minipage}{0.32\linewidth}
    \vspace{-2em}
    \centering
    \includegraphics[width=\linewidth]{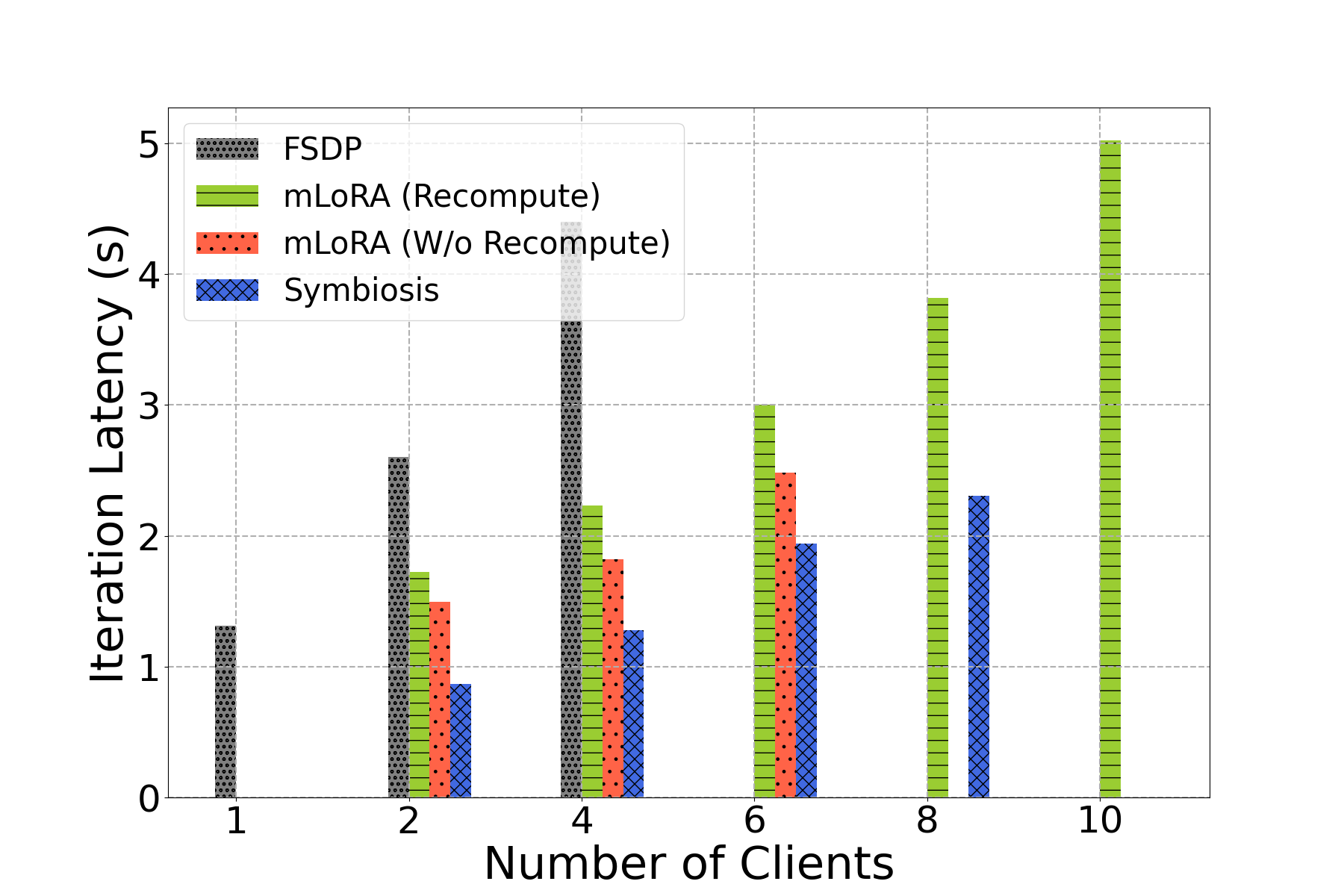}
    \caption{\footnotesize {\bf Sharded local:} Iteration latency with Llama2-13B, batch Size=2, sequence length=512}
    \label{fig:ft-fsdp-local-13b}
    \end{minipage}\hfill
    \begin{minipage}{0.32\linewidth}
    \vspace{-2em}
    \centering
    \includegraphics[width=\linewidth]{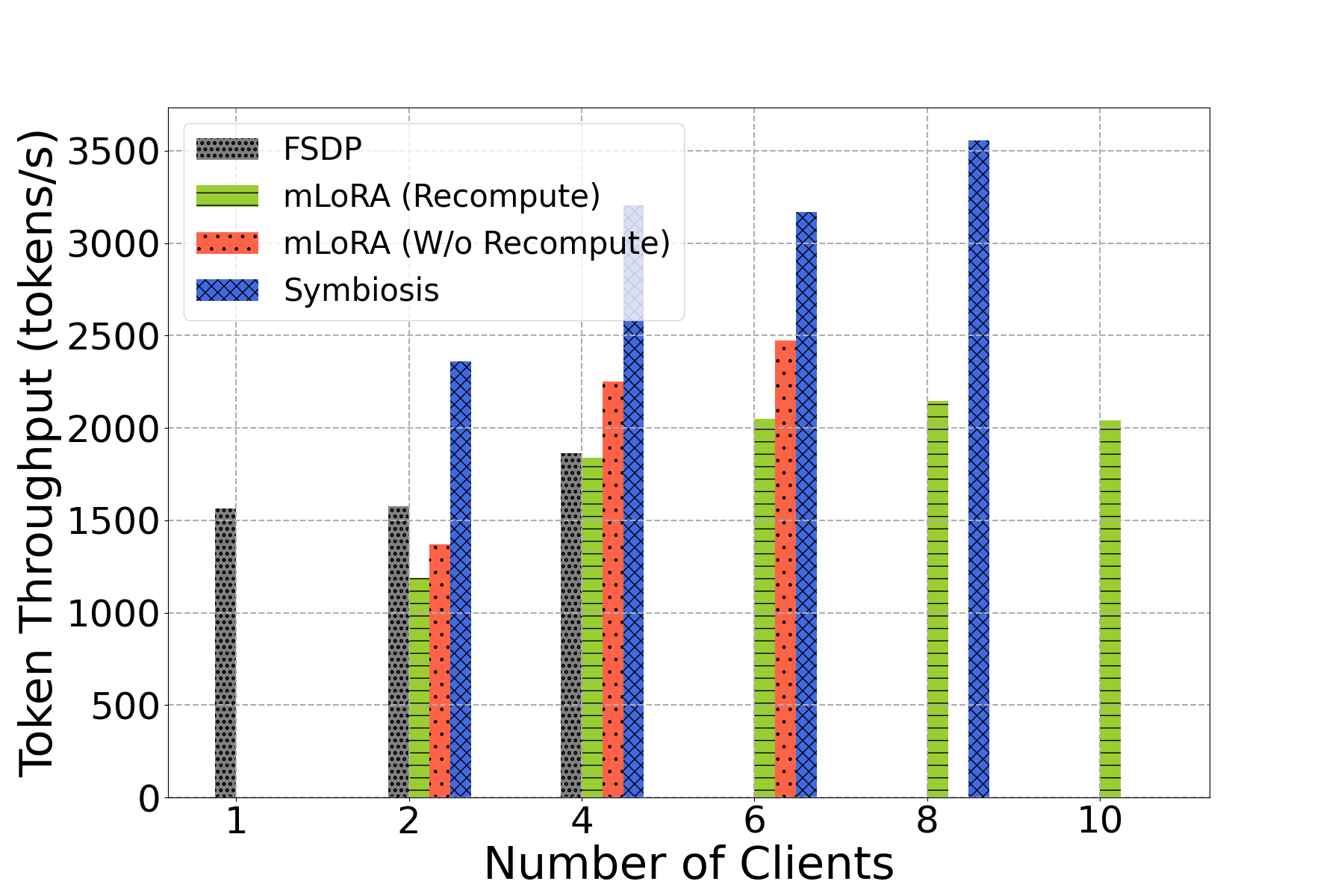}
    \caption{\footnotesize {\bf Sharded local:} Token Throughput with Llama2-13B, batch size=2, sequence length=512.}
    \label{fig:ft-fsdp-local-13b-thr}
    \end{minipage}\hfill
    \begin{minipage}{0.32\linewidth}
    \vspace{-0.5em}
    \centering
    \includegraphics[width=\linewidth]{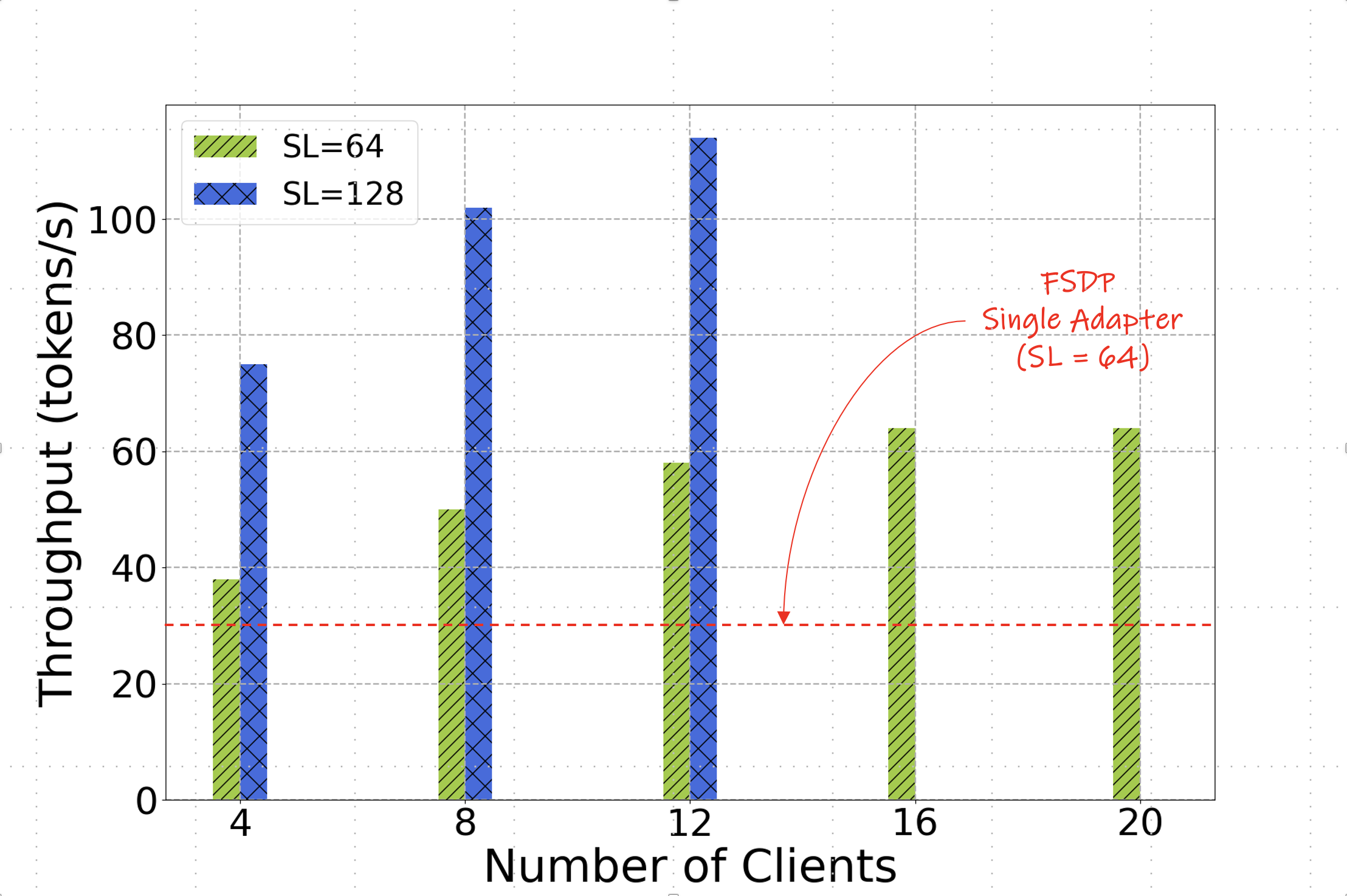}
    \caption{\footnotesize {\bf Sharded Remote:} Throughput for Gemma2-27B, batch size=2. The base executor is sharded across 4 GPUs, the clients are hosted on other 4 GPUs.}
    \label{fig:ft-fsdp-4-gpu}
    \end{minipage}\hfill
    \caption*{}
\end{figure*}
\begin{figure*}[h]
    \begin{minipage}{0.31\linewidth}
    \centering
    \includegraphics[width=\linewidth]{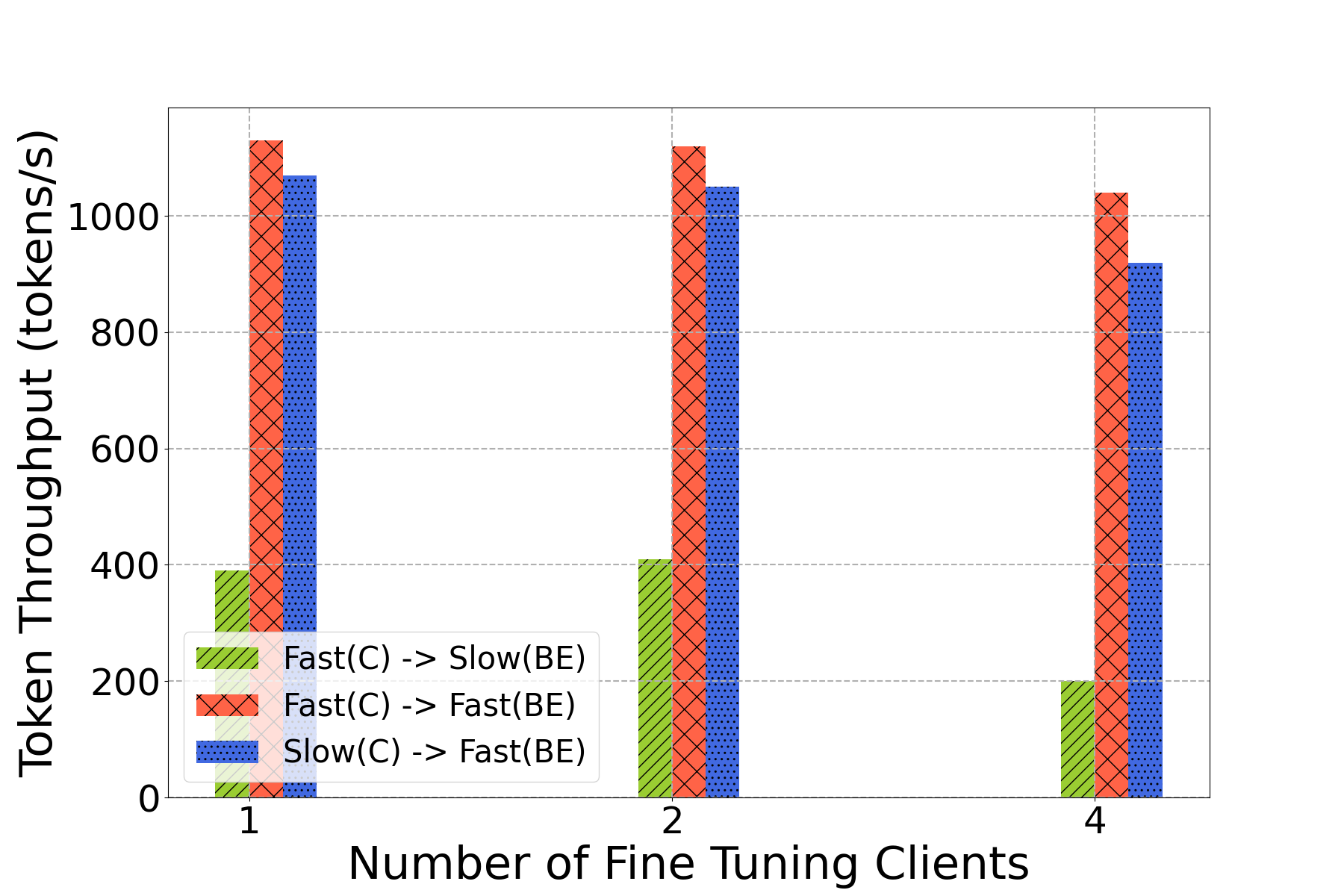}
    \caption{\footnotesize {\bf Heterogeneous GPUs:} Throughput for Llama2-13B, batch size=2. C - Client, B - Base executor. Fast represents 350W GPU, Slow represents 100W GPU.}
    \label{fig:ft-hg-2gpu}
    \end{minipage}\hfill 
    \begin{minipage}{0.32\linewidth}
    \centering
    \includegraphics[width=\linewidth]{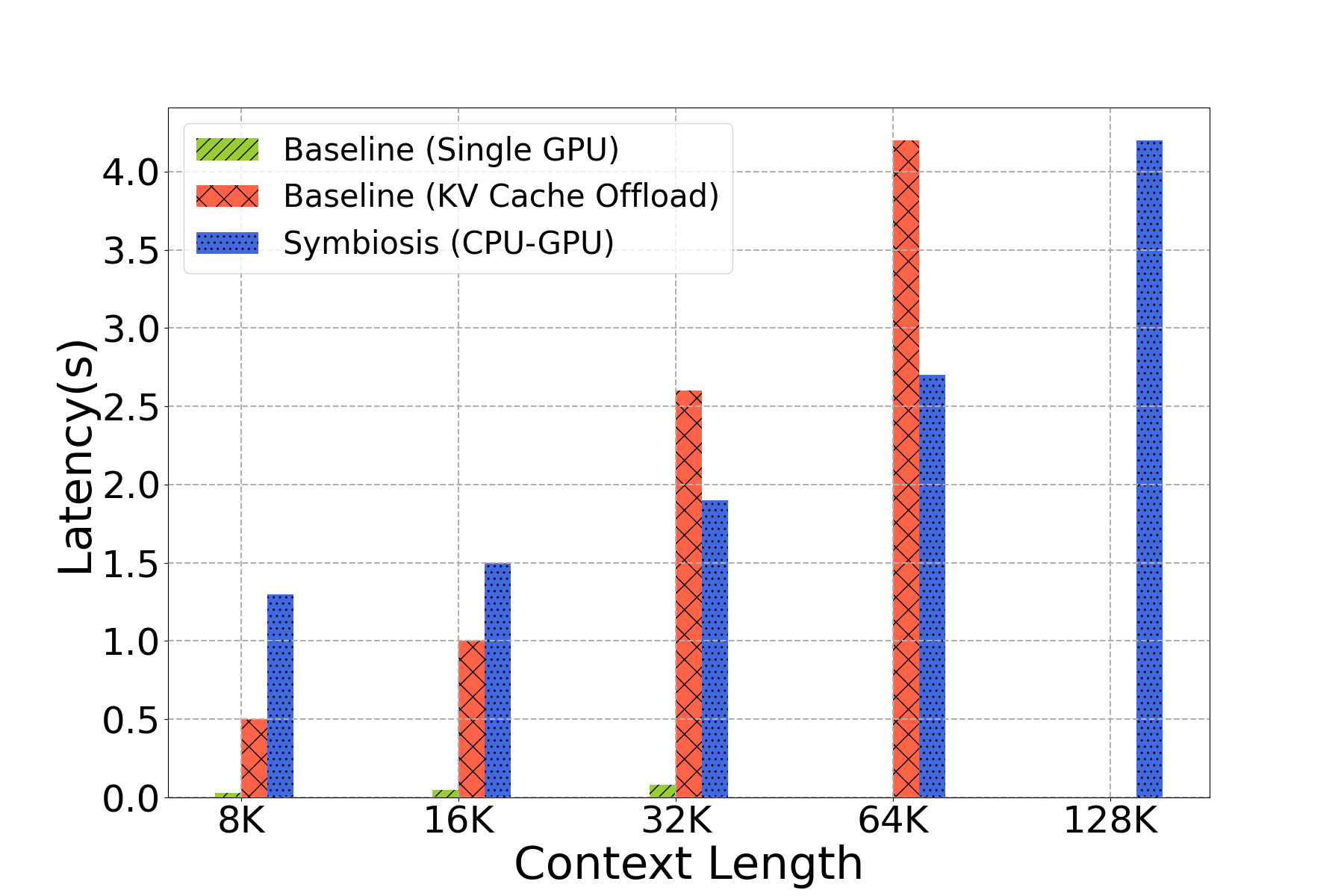}
    \caption{\footnotesize {\bf CPU-GPU Inference:} KV Cache size is proportional to the Context (Sequence Length). E.g., 128K context length = 64GB KV Cache for Llama2-7B.}
    \label{fig:cpu-gpu-hg-inf}
    \end{minipage}\hfill
    \begin{minipage}{0.32\linewidth}
    \centering
    \includegraphics[width=\linewidth]{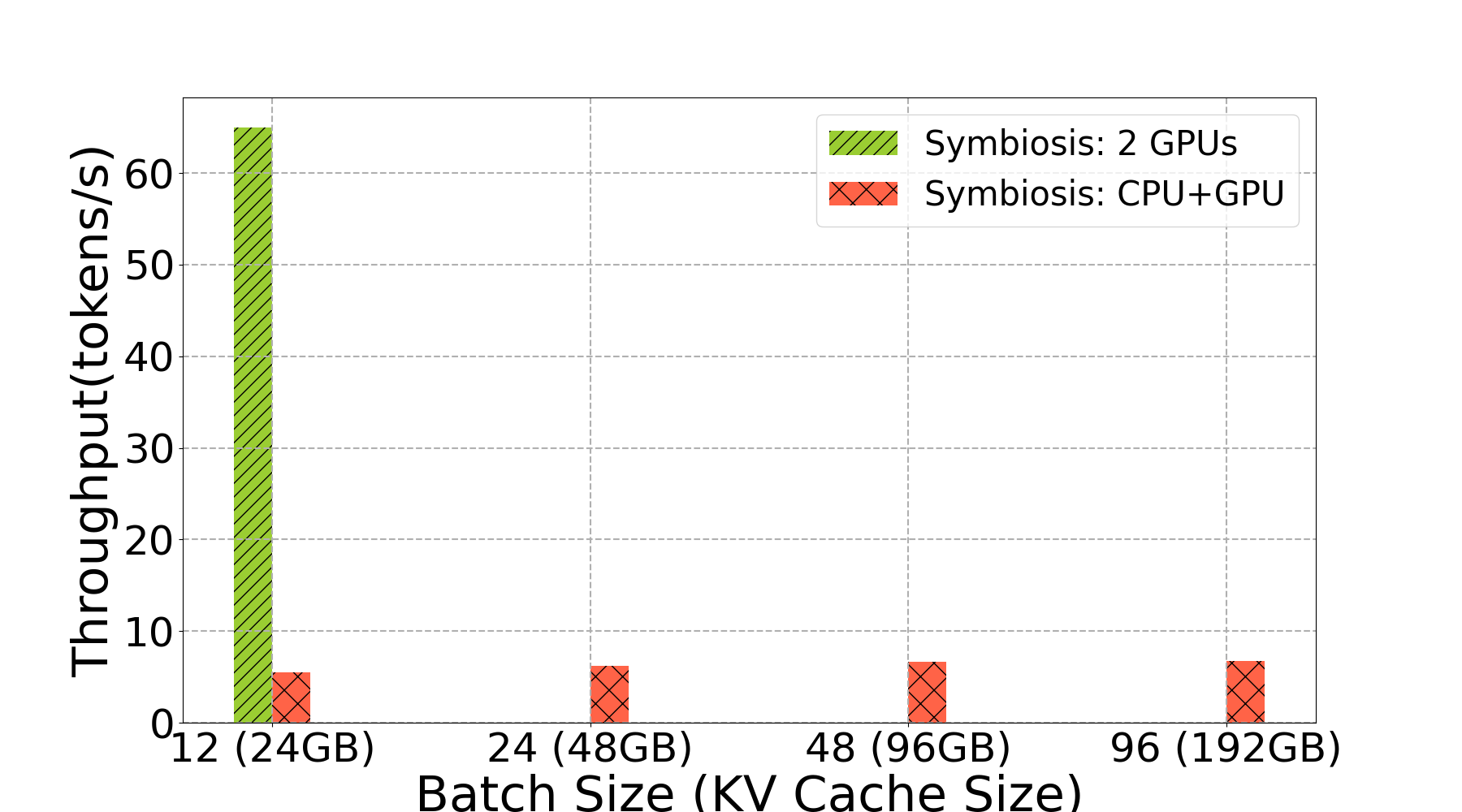}
    \caption{\footnotesize {\bf CPU-GPU Inference with Multiple Requests:} Model=Llama2-7b. The client is located on CPU and the base executor is located on GPU.}
    \label{fig:cpu-gpu-inf}
    \end{minipage}\hfill
\end{figure*}

\paragraph{Sharded Local}
In sharded local configuration, we run both base executor and clients on same set of GPUs. Sharded mode spans the base model across 2 GPUs, whereas a client can execute on any one of the GPUs. Figure~\ref{fig:ft-fsdp-local-13b} shows the latency comparison with mLoRA for Llama2-13B model. mLoRA can either optimize the memory consumption with recompute while sacrificing performance, or it can achieve better performance with higher memory consumption, and as a result, it accommodates fewer adapters before running out of memory. In contrast, because of the optimized backward pass, \scheme{} is both memory and performance optimized. Therefore, it it able to run more fine-tuning clients while achieving lower latency and higher throughput (Figure~\ref{fig:ft-fsdp-local-13b-thr}).

We also compare the performance of \scheme{} with FSDP baseline with Llama2-13B model. The iteration latency with \scheme{} is almost 2X lower than that of FSDP (considering it processes two tokens in an iteration). This is because FDSP shards exchanges gradients to train a common adapter. Moreover, optimized backward pass helps further improve the throughput and reduce the iteration latency.

When measured in terms of memory, for Llama2-13B model, FSDP occupies 17GB of memory on each of the two GPUs. For comparison, this means that 4 FSDP processes can be run in parallel on 2 GPUs (as shown in Figure~\ref{fig:ft-fsdp-local-13b-thr}) to fine-tune 4 adapters. In contrast, \scheme{} can fine-tune 4 adapters in almost half the time.

\paragraph{Sharded Remote}
In sharded remote configuration, we serve the base model across 4 GPUs and distribute clients across separate set of 4 GPUs. This configuration is best suited for large models that cannot be accommodated on a single GPU to serve memory intensive clients that cannot be co-located with the base model.

Figure~\ref{fig:ft-fsdp-4-gpu} shows the throughput with increasing clients for Gemma2-27B model. For comparison, the baseline FSDP over 8 GPUs fine-tunes a single adapter at 32 tokens/s with batch size of 2 and sequence length of 64. The primary source of overhead with both baseline and \scheme{} is from parameter fetching. Additionally, the baseline needs to exchange gradients for fine-tuning a common adapter. As a result, \scheme{} with 8 adapters (clients) outperforms a comparable single adapter 8-GPU FSDP baseline, where each shard processes separate tokens.

\subsection{\scheme{} with Heterogeneous Resources}
In this section, we show the deployment of \scheme{} across heterogeneous GPUs and across GPU-CPU. We show that \scheme{} can make a better use of heterogeneous resources by decoupling the compute-light and memory-bound clients from compute-heavy base model and offloading them on less powerful accelerators or memory abundant CPUs.

\subsubsection{Heterogeneous GPU Fine-Tuning}
For multi-GPU configuration, we use a combination of less powerful (100W) and more powerful (350W) GPUs. Also, GPUs used for this experiment only have 40GB of memory. Figure~\ref{fig:ft-hg-2gpu} shows the fine-tuning throughput with increasing clients for Llama2-13B model. With \scheme{}, only the powerful GPU serves the more compute intensive base model layers, whereas the less powerful GPU incorporates the less compute intensive adapter fine-tuning and attention. Therefore, the heterogeneous setup has little impact on the fine-tuning performance and performs equally well as hosting both on faster GPUs.

\begin{figure*}[h]
    \begin{minipage}{0.32\linewidth}
    \vspace{-1em}
    \centering
    \includegraphics[width=\linewidth]{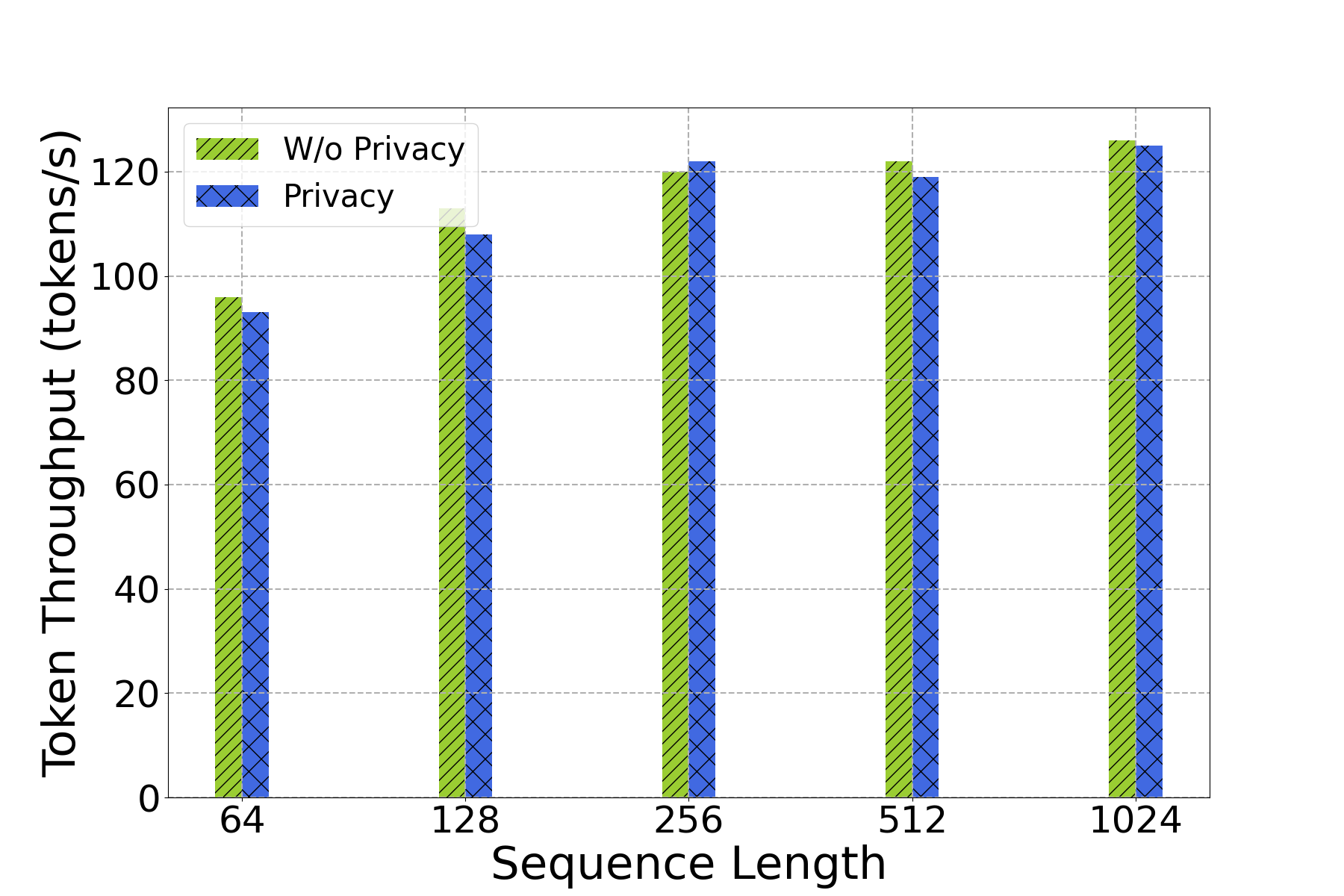}
    \caption{\footnotesize Privacy has little impact on inference performance in over-the-network setup.}
    \label{fig:privacy}
    \end{minipage}\hfill
    \begin{minipage}{0.32\linewidth}
    \centering
    \includegraphics[width=\linewidth]{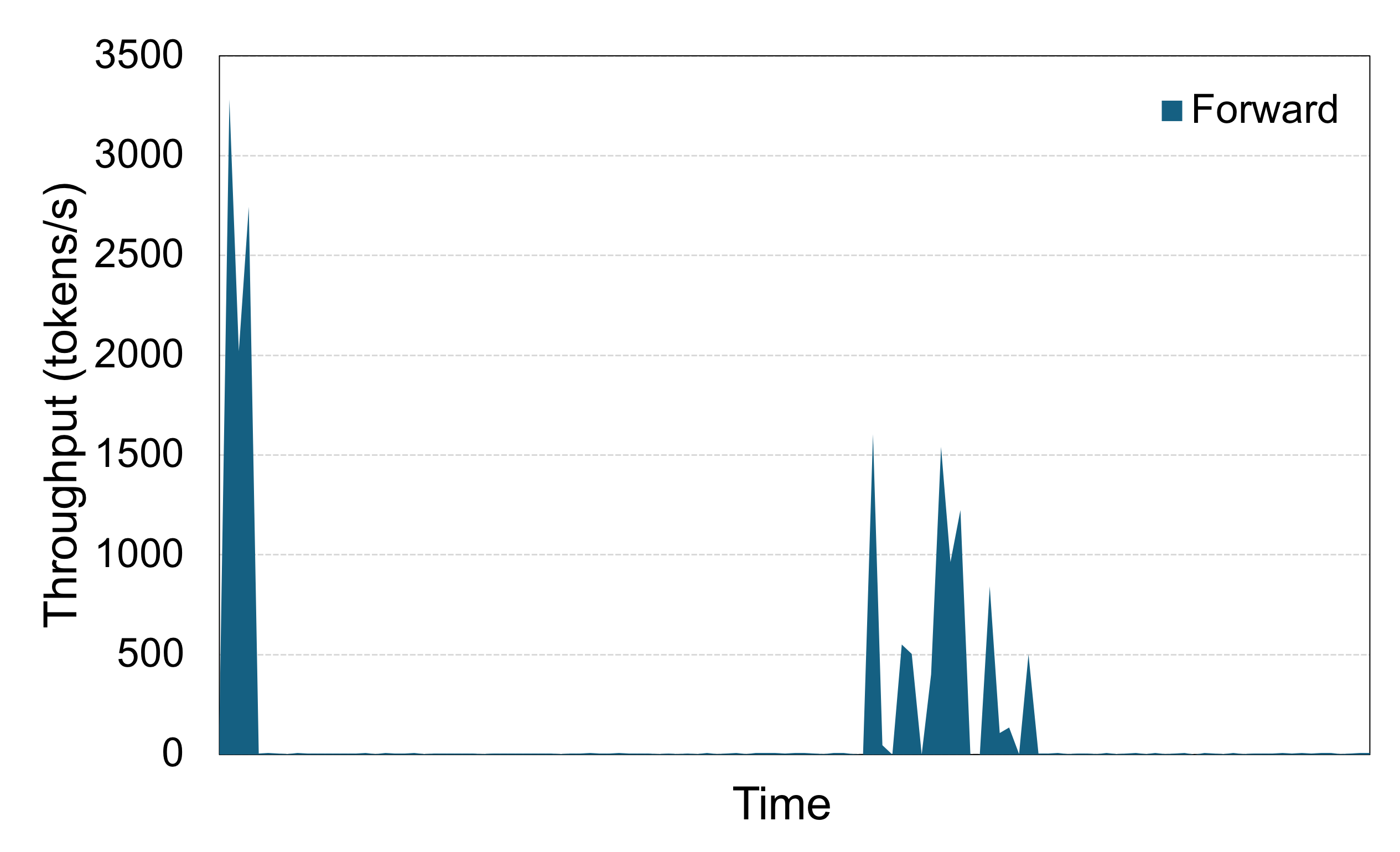}
    \caption{\footnotesize Token generation throughput 8 inference clients for Llama2-7B, batch size=2 and sequence length of 512.}
    \label{fig:mixed-inf}
    \end{minipage}\hfill
    \begin{minipage}{0.32\linewidth}
    \centering
    \includegraphics[width=\linewidth]{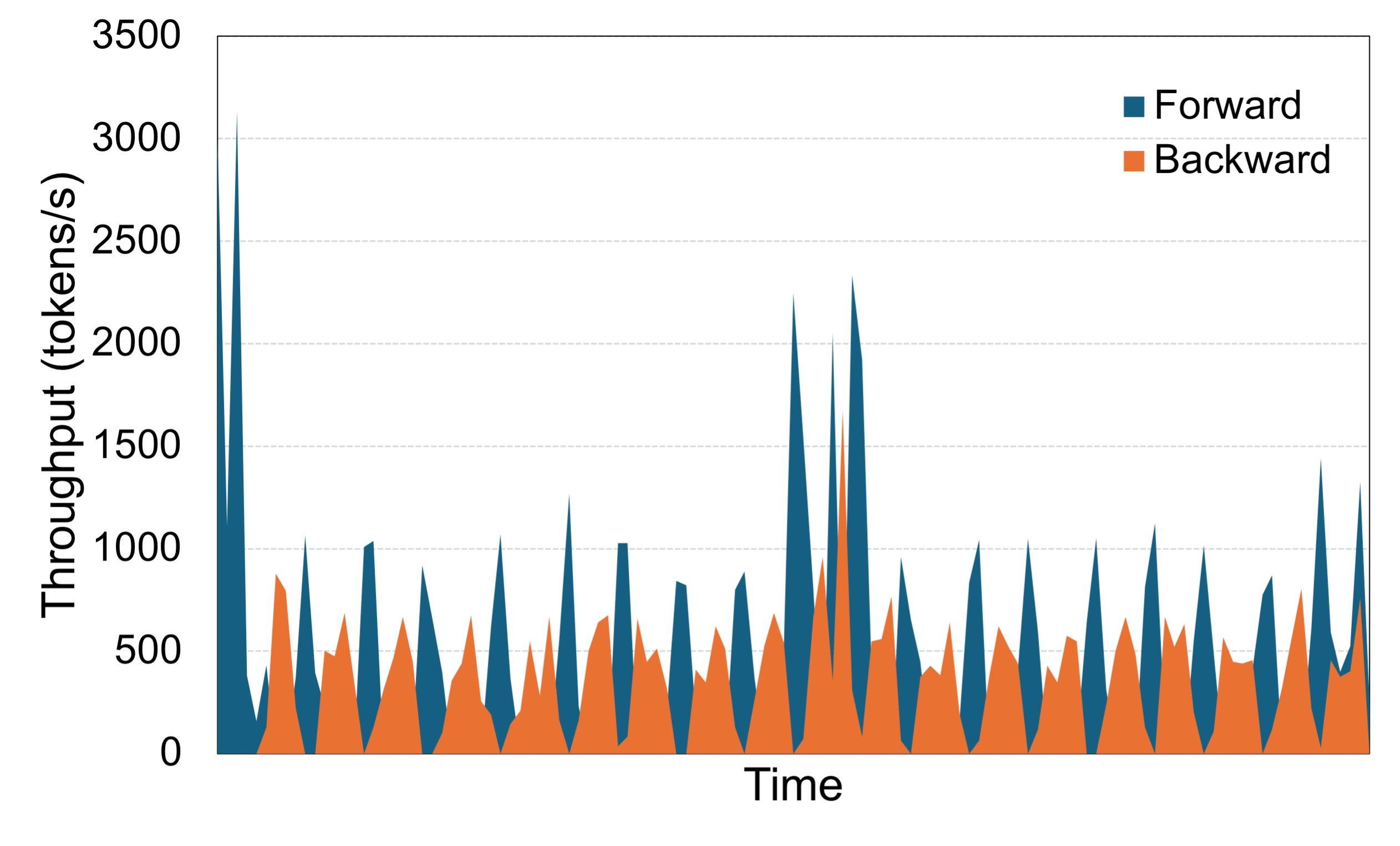}
    \caption{\footnotesize Token generation throughput 6 inference clients and 2 fine-tuning clients for Llama2-7B, batch size=2 and sequence length of 512.}
    \label{fig:mixed-inf-ft}
    \end{minipage}\hfill
\end{figure*}

\subsubsection{Inference with Heterogeneous Client}
\label{eval:cpu-gpu-hg-inf}
\paragraph{Single Request:}
Figure~\ref{fig:cpu-gpu-hg-inf} shows the inter-token latency of inference with the system proposed in Section~\ref{sec:hg-inf} with varying context length. Here, the client executes the prefill stage on GPU with a CPU-offloaded KV cache and the decoding stage on CPU.
We compare our results with two GPU-based baselines. As expected, the first baseline with both cache and the corresponding compute on the GPU is the fastest. However, in our experiments with Llama2-7B, it fails for KV cache sizes greater than 16GB due to the limited GPU memory. The second baseline with KV cache offloaded to CPU but compute still performed on GPU is faster than \scheme{} initially but slows down with the increasing context length. For the context length of 32K and beyond in Llama2-7B, the performance cost of transferring offloaded KV cache tensors from CPU to GPU exceeds the acceleration benefits provided by GPU. In contrast, \scheme{} has constant CPU-GPU data transfer overhead irrespective of the KV cache size. The increase in latency is from CPU based computation of attention for increasing sequence length. This trade-off breaks in favor of \scheme{}. As a result, we see that \scheme{}'s heterogeneous inference client is 33\% faster than the GPU baseline. Moreover, the second baseline also cannot accommodate a fraction of KV cache with larger sequence lengths, thus runs out of memory sooner. Whereas, \scheme{} is able to support longer context length.

\paragraph{Multiple Requests:}
Figure~\ref{fig:cpu-gpu-inf} shows the inference throughput with multiple batched requests, each with 1K sequence length. With 2 GPUs, the client and base executor are placed on different GPUs. However, the 40GB GPU used here for the client cannot accommodate the KV cache for 24 or more requests. In comparison, even though the CPU-side client suffers from higher request latency, it can accommodate 8X as many requests at 7.5 tokens/s throughput. 


\begin{table}[t]
    \centering
    \begin{tabular}{p{3.0cm}P{2.5cm}}
        \toprule
         Configuration & Latency\\
        \bottomrule
         Small \& Small & 0.30\\
        \hline
         Small \& Large & 3.74\\
        \hline
         Large \& Large & 6.94\\
        \hline
    \end{tabular}
    \vspace{1em}
    \caption{\footnotesize vLLM (v0.7.0) prefill response time for batched inference of small and large requests for Llama2-7B model. Small: sequence length=1, Large: sequence length=512.}
    \label{tab:vllm}
\end{table}

\begin{table*}[t]
    \centering
    \begin{tabular}{p{3.2cm}|P{1.8cm}P{1.8cm}|P{1.8cm}P{1.8cm}|P{1.8cm}P{1.8cm}}
        \toprule
         & & & \multicolumn{2}{c|}{\bf{Prefill}} & \multicolumn{2}{c}{\bf{Generation}} \\
         & Request Rate & Average & Throughput & Latency & Throughput & Latency \\
         & (reqs/s) & Batch Size & (tokens/s) & (s) & (tokens/s) & (s) \\
        \bottomrule
         No Lockstep & 0.72 & 1.1 & 18912 & 8.7 & 13.7 & 2.3 \\
        \hline
         Opportunistic Batching & 0.87 & 2.14 & 20392 & 7.2 & 17.6 & 2 \\
        \hline
    \end{tabular}
    \vspace{1em}
    \caption{\footnotesize Average latency and aggregate throughput of 32 inference jobs in 8-GPU sharded local configuration of Llama2-13B. Opportunistic batching achieves 20\% higher request rate. Moreover, it improves token throughput while lowering request latency.}
    \vspace{-1em}
    \label{tab:lockstep}

\end{table*}

\subsection{Mixed Fine-tuning and Inference}
In this section, we show the use of \scheme{} to share a model between inference and fine-tuning jobs. We demonstrate that having a common platform allows service provider to improve GPU utilization by time multiplexing inference and fine-tuning jobs, i.e., when there are not enough inference or fine-tuning requests. Moreover, batching of inference and fine tuning requests further improves utilization.

Figure~\ref{fig:mixed-inf} shows the combined throughput of 8 inference jobs. The base executor and inference clients are hosted on separate GPUs.
The inference requests consists of prefill and generation phases. It can be observed that while the throughput with the prefill is visibly higher, the GPU remains under utilized during the generation phase. This is because with the batch size of 2, the GPU processes only two tokens at a time for a given client. \scheme{} improves the utilization of GPUs by also incorporating the fine-tuning clients. In Figure~\ref{fig:mixed-inf-ft}, we replace 2 of the inference clients with fine-tuning clients. The long sequence and backward pass of fine-tuning clients improve the system throughput.

However, mixing of inference and fine-tuning jobs may result in degradation of inference requests from increased response time. In a variety of use cases, the inference jobs directly serve end-users, thus providing quicker response is important to preserve the quality of service. To accomplish this, on a shared platform, \scheme{} prioritizes the inference requests through opportunistic batching, i.e., when possible, at each layer, inference requests are batched with other inference or fine-tuning requests. This approach preserves the low response time requirement of inference while improving the system utilization through better batching. Therefore, the token generation latency of inference requests for inference only and mixed workloads remains roughly the same at 1.4s.

\subsection{Opportunistic Batching}
Table~\ref{tab:vllm} shows lockstep execution of two inference requests executed in the same batch in vLLM. Both requests use adapter LoRA 3 from Table~\ref{tab:lora}. It can be observed that when batching large and small requests together, because of the lockstep execution, the response time of the small request also suffers.

In comparison, Table~\ref{tab:lockstep} shows the benefit of opportunistic batching for inference workload from Azure Public Dataset~\cite{azure-dataset}. 
This is a sample of the traces from multiple LLM inference services in Azure. The dataset contains a trace of inference requests with context sequence lengths of up to 7K tokens and generated sequence lengths of up to 32 tokens. We use Llama2-13B as the base model deployed in sharded local configuration across 8 GPUs. 32 inference clients share the base model for inference (4 clients per GPU). Each client uses a 64-rank LoRA adapter with q and k as the fine-tuned layers. Together all clients replay a portion of the inference workload trace from the Azure Public Dataset.
In the {\em no-lockstep} approach, each request can progress through the model as quickly as possible without having to batch with other requests. However, with a large number of requests, such approach causes the execution of different layers by different requests to be serialized. The serialization of requests increases latency and the lack of batching reduces inference throughput. In contrast, \scheme{}'s {\em opportunistic batching} waits only for a pre-determined duration for batching and then proceeds with the accumulated requests. This improves throughput from better batching. For instance, the generation throughput with opportunistic batching is 28\% higher than that with no-lockstep. Note that the throughput improvement in prefill is not as significant as generation from the already large context lengths, which limit the further throughput improvement from batching. Additional benefit of opportunistic batching is that the requests batched at the first layer are not required to be batched again for the following layers, which makes different rate of execution for different clients possible. This behavior is useful in maintaining lower latency.

\subsection{Privacy for Multi-Tenancy}
Figure~\ref{fig:privacy} demonstrates the effect of privacy on inference performance of model Llama2-7B. We host a client on a separate host, with the assumption that such host is under tenant control. Whereas the base model can be hosted by a service provider. The tenant adds noise to the activation before transmitting them over the network and the noise effect is deducted from the received output. It can be observed that the effect of noise addition and subtraction is minimal. This is because we calculate the noise effect for each layer in advance. Moreover, the network interference is a primary factor in the degraded performance when compared to the communication between the GPUs on the same host. However, even with the multi-tenant setup the performance remains acceptable. \scheme{} uniquely enables this use case where tenant parameters (of the adapter) and activations are protected, while allowing them to leverage a common base model.
\section{Conclusions}
In \scheme{}, we present an inference and fine-tuning platform. \scheme{} provides an abstraction of a base model as-a-service thus enabling flexible placement, batching and optimizations at a granularity of a model layer. We leverage the abstraction to decouple the client specific state, adapters from the base model. This decoupling enables new use cases, namely, (a) base model as-a-service where a shared base model can serve different clients, (b) Inference and fine-tuning over heterogeneous resources, (c) Privacy preserving multi-tenant platform. Moreover, \scheme{} works out-of-the-box for variety of models in transformer library without requiring any changes to the model code.
\section*{Acknowledgment}
Thanks to {\bf Ka-Ho Chow} for his contribution to the initial prototype, to {\bf Ryan Marcus} for his guidance as a shepherd, and to our reviewers for their valuable suggestions. 
\bibliography{main}


\begin{thebibliography}{54}


\ifx \showCODEN    \undefined \def \showCODEN     #1{\unskip}     \fi
\ifx \showDOI      \undefined \def \showDOI       #1{#1}\fi
\ifx \showISBNx    \undefined \def \showISBNx     #1{\unskip}     \fi
\ifx \showISBNxiii \undefined \def \showISBNxiii  #1{\unskip}     \fi
\ifx \showISSN     \undefined \def \showISSN      #1{\unskip}     \fi
\ifx \showLCCN     \undefined \def \showLCCN      #1{\unskip}     \fi
\ifx \shownote     \undefined \def \shownote      #1{#1}          \fi
\ifx \showarticletitle \undefined \def \showarticletitle #1{#1}   \fi
\ifx \showURL      \undefined \def \showURL       {\relax}        \fi
\providecommand\bibfield[2]{#2}
\providecommand\bibinfo[2]{#2}
\providecommand\natexlab[1]{#1}
\providecommand\showeprint[2][]{arXiv:#2}

\bibitem[off(2024)]%
        {offloadedcache}
 \bibinfo{year}{2024}\natexlab{}.
\newblock \bibinfo{title}{Best Practices for Generation with Cache}.
\newblock
  \bibinfo{howpublished}{\url{https://huggingface.co/docs/transformers/en/kv_cache\#offloaded-cache}}.
\newblock
\newblock
\shownote{Accessed: 2024-09-23}.


\bibitem[azu(2025)]%
        {azure-dataset}
 \bibinfo{year}{2025}\natexlab{}.
\newblock \bibinfo{title}{Azure LLM inference Trace 2024}.
\newblock
\newblock
\urldef\tempurl%
\url{https://github.com/Azure/AzurePublicDataset/blob/master/AzureLLMInferenceDataset2024.md}
\showURL{%
\tempurl}


\bibitem[aws(2025)]%
        {aws-bs}
 \bibinfo{year}{2025}\natexlab{}.
\newblock \bibinfo{title}{Fine-tune Meta Llama 3.2 using Amazon SageMaker}.
\newblock
\newblock
\urldef\tempurl%
\url{https://aws.amazon.com/blogs/machine-learning/fine-tune-meta-llama-3-2-text-generation-models-for-generative-ai-inference-using-amazon-sagemaker-jumpstart}
\showURL{%
\tempurl}


\bibitem[fle(2025)]%
        {flexflow}
 \bibinfo{year}{2025}\natexlab{}.
\newblock \bibinfo{title}{FlexFlow: DNN Framework}.
\newblock
\newblock
\urldef\tempurl%
\url{"https://flexflow.ai"}
\showURL{%
\tempurl}


\bibitem[ibm(2025)]%
        {ibm-bs}
 \bibinfo{year}{2025}\natexlab{}.
\newblock \bibinfo{title}{LoRA fine-tuning Granite LLM}.
\newblock
\newblock
\urldef\tempurl%
\url{https://www.ibm.com/think/tutorials/lora-fine-tuning-granite-llm}
\showURL{%
\tempurl}


\bibitem[uns(2025)]%
        {unsloth-bs}
 \bibinfo{year}{2025}\natexlab{}.
\newblock \bibinfo{title}{LoRA Hyperparameters Guide}.
\newblock
\newblock
\urldef\tempurl%
\url{https://docs.unsloth.ai/get-started/fine-tuning-llms-guide/lora-hyperparameters-guide}
\showURL{%
\tempurl}


\bibitem[mig(2025)]%
        {mig}
 \bibinfo{year}{2025}\natexlab{}.
\newblock \bibinfo{title}{NVIDIA H100 MIG Documentation}.
\newblock
  \bibinfo{howpublished}{\url{https://docs.nvidia.com/launchpad/ai/h100-mig/latest/h100-mig-gpu.html}}.
\newblock
\newblock
\shownote{Accessed: 2025-01-14}.


\bibitem[Agarwal et~al\mbox{.}(2024)]%
        {symphony}
\bibfield{author}{\bibinfo{person}{Saurabh Agarwal}, \bibinfo{person}{Anyong
  Mao}, \bibinfo{person}{Aditya Akella}, {and} \bibinfo{person}{Shivaram
  Venkataraman}.} \bibinfo{year}{2024}\natexlab{}.
\newblock \bibinfo{title}{SYMPHONY: Improving Memory Management for LLM
  Inference Workloads}.
\newblock
\newblock
\showeprint[arxiv]{2412.16434}~[cs.DC]
\urldef\tempurl%
\url{https://arxiv.org/abs/2412.16434}
\showURL{%
\tempurl}


\bibitem[Agrawal et~al\mbox{.}(2023)]%
        {sarathi}
\bibfield{author}{\bibinfo{person}{Aayush Agrawal}, \bibinfo{person}{Animesh
  Panwar}, \bibinfo{person}{Jatin Mohan}, \bibinfo{person}{Naman Kwatra},
  \bibinfo{person}{Bhargav~S. Gulavani}, {and} \bibinfo{person}{Ramachandran
  Ramjee}.} \bibinfo{year}{2023}\natexlab{}.
\newblock \showarticletitle{SARATHI: Efficient LLM Inference by Piggybacking
  Decodes with Chunked Prefills}.
\newblock \bibinfo{journal}{\emph{arXiv preprint arXiv:2308.16369}}
  (\bibinfo{year}{2023}).
\newblock
\urldef\tempurl%
\url{https://arxiv.org/abs/2308.16369}
\showURL{%
\tempurl}


\bibitem[Ahmed et~al\mbox{.}(2024)]%
        {ft-precision}
\bibfield{author}{\bibinfo{person}{Syed Ahmed}, \bibinfo{person}{Christian
  Sarofeen}, \bibinfo{person}{Mike Ruberry}, \bibinfo{person}{Eddie Yan},
  \bibinfo{person}{Natalia Gimelshein}, \bibinfo{person}{Michael Carilli},
  \bibinfo{person}{Szymon Migacz}, \bibinfo{person}{Piotr Bialecki},
  \bibinfo{person}{Paulius Micikevicius}, \bibinfo{person}{Dusan Stosic},
  \bibinfo{person}{Dong Yang}, {and} \bibinfo{person}{Naoya Maruyama}.}
  \bibinfo{year}{2024}\natexlab{}.
\newblock \showarticletitle{What Every User Should Know About Mixed Precision
  Training in PyTorch}.
\newblock  (\bibinfo{year}{2024}).
\newblock
\urldef\tempurl%
\url{https://pytorch.org/blog/what-every-user-should-know-about-mixed-precision-training-in-pytorch/}
\showURL{%
\tempurl}


\bibitem[Aminabadi et~al\mbox{.}(2022)]%
        {deepspeed}
\bibfield{author}{\bibinfo{person}{Reza~Yazdani Aminabadi},
  \bibinfo{person}{Samyam Rajbhandari}, \bibinfo{person}{Ammar~Ahmad Awan},
  \bibinfo{person}{Cheng Li}, \bibinfo{person}{Du Li}, \bibinfo{person}{Elton
  Zheng}, \bibinfo{person}{Olatunji Ruwase}, \bibinfo{person}{Shaden Smith},
  \bibinfo{person}{Minjia Zhang}, \bibinfo{person}{Jeff Rasley},
  {et~al\mbox{.}}} \bibinfo{year}{2022}\natexlab{}.
\newblock \showarticletitle{Deepspeed-inference: enabling efficient inference
  of transformer models at unprecedented scale}. In
  \bibinfo{booktitle}{\emph{Proceedings of the International Conference on High
  Performance Computing, Networking, Storage and Analysis (SC)}}.
\newblock


\bibitem[Chen et~al\mbox{.}(2023)]%
        {punica}
\bibfield{author}{\bibinfo{person}{Lequn Chen}, \bibinfo{person}{Zihao Ye},
  \bibinfo{person}{Yongji Wu}, \bibinfo{person}{Danyang Zhuo},
  \bibinfo{person}{Luis Ceze}, {and} \bibinfo{person}{Arvind Krishnamurthy}.}
  \bibinfo{year}{2023}\natexlab{}.
\newblock \bibinfo{title}{Punica: Multi-Tenant LoRA Serving}.
\newblock
\newblock
\showeprint[arxiv]{2310.18547}~[cs.DC]
\urldef\tempurl%
\url{https://arxiv.org/abs/2310.18547}
\showURL{%
\tempurl}


\bibitem[Chen et~al\mbox{.}(2024)]%
        {lamina}
\bibfield{author}{\bibinfo{person}{Shaoyuan Chen}, \bibinfo{person}{Yutong
  Lin}, \bibinfo{person}{Mingxing Zhang}, {and} \bibinfo{person}{Yongwei Wu}.}
  \bibinfo{year}{2024}\natexlab{}.
\newblock \showarticletitle{Efficient and Economic Large Language Model
  Inference with Attention Offloading}.
\newblock \bibinfo{journal}{\emph{arXiv preprint arXiv:2405.01814}}
  (\bibinfo{year}{2024}).
\newblock


\bibitem[Contributors(2023)]%
        {lmdeploy2023}
\bibfield{author}{\bibinfo{person}{LMDeploy Contributors}.}
  \bibinfo{year}{2023}\natexlab{}.
\newblock \bibinfo{title}{LMDeploy: A Toolkit for Compressing, Deploying, and
  Serving LLM}.
\newblock \bibinfo{howpublished}{\url{https://github.com/InternLM/lmdeploy}}.
\newblock


\bibitem[Contributors(2024)]%
        {lorax2024}
\bibfield{author}{\bibinfo{person}{LoRAX Contributors}.}
  \bibinfo{year}{2024}\natexlab{}.
\newblock \bibinfo{title}{LoRAX: Multi-LoRA inference server that scales to
  1000s of fine-tuned LLMs}.
\newblock \bibinfo{howpublished}{\url{https://github.com/lorax/lorax}}.
\newblock
\newblock
\shownote{Accessed: 2024-09-16}.


\bibitem[Fu et~al\mbox{.}(2024)]%
        {llm-sched}
\bibfield{author}{\bibinfo{person}{Yichao Fu}, \bibinfo{person}{Siqi Zhu},
  \bibinfo{person}{Runlong Su}, \bibinfo{person}{Aurick Qiao},
  \bibinfo{person}{Ion Stoica}, {and} \bibinfo{person}{Hao Zhang}.}
  \bibinfo{year}{2024}\natexlab{}.
\newblock \bibinfo{title}{Efficient LLM Scheduling by Learning to Rank}.
\newblock
\newblock
\showeprint[arxiv]{2408.15792}~[cs.LG]
\urldef\tempurl%
\url{https://arxiv.org/abs/2408.15792}
\showURL{%
\tempurl}


\bibitem[Gao et~al\mbox{.}(2024)]%
        {cachedattention}
\bibfield{author}{\bibinfo{person}{Bin Gao}, \bibinfo{person}{Zhuomin He},
  \bibinfo{person}{Puru Sharma}, \bibinfo{person}{Qingxuan Kang},
  \bibinfo{person}{Djordje Jevdjic}, \bibinfo{person}{Junbo Deng},
  \bibinfo{person}{Xingkun Yang}, \bibinfo{person}{Zhou Yu}, {and}
  \bibinfo{person}{Pengfei Zuo}.} \bibinfo{year}{2024}\natexlab{}.
\newblock \showarticletitle{Cost-Efficient Large Language Model Serving for
  Multi-turn Conversations with CachedAttention}. In
  \bibinfo{booktitle}{\emph{Proceedings of the 2024 USENIX Annual Technical
  Conference}}.
\newblock


\bibitem[Gao et~al\mbox{.}(2018)]%
        {cellbatch}
\bibfield{author}{\bibinfo{person}{Pin Gao}, \bibinfo{person}{Lingfan Yu},
  \bibinfo{person}{Yongwei Wu}, {and} \bibinfo{person}{Jinyang Li}.}
  \bibinfo{year}{2018}\natexlab{}.
\newblock \showarticletitle{Low latency RNN inference with cellular batching}.
  In \bibinfo{booktitle}{\emph{Proceedings of the Thirteenth EuroSys
  Conference}} (Porto, Portugal) \emph{(\bibinfo{series}{EuroSys '18})}.
  \bibinfo{publisher}{Association for Computing Machinery},
  \bibinfo{address}{New York, NY, USA}, Article \bibinfo{articleno}{31},
  \bibinfo{numpages}{15}~pages.
\newblock
\showISBNx{9781450355841}
\urldef\tempurl%
\url{https://doi.org/10.1145/3190508.3190541}
\showDOI{\tempurl}


\bibitem[Gong et~al\mbox{.}(2020)]%
        {me-attack2}
\bibfield{author}{\bibinfo{person}{Xueluan Gong}, \bibinfo{person}{Qian Wang},
  \bibinfo{person}{Yanjiao Chen}, \bibinfo{person}{Wang Yang}, {and}
  \bibinfo{person}{Xinchang Jiang}.} \bibinfo{year}{2020}\natexlab{}.
\newblock \showarticletitle{Model Extraction Attacks and Defenses on
  Cloud-Based Machine Learning Models}.
\newblock \bibinfo{journal}{\emph{IEEE Communications Magazine}}
  \bibinfo{volume}{58}, \bibinfo{number}{12} (\bibinfo{year}{2020}),
  \bibinfo{pages}{83--89}.
\newblock
\urldef\tempurl%
\url{https://doi.org/10.1109/MCOM.001.2000196}
\showDOI{\tempurl}


\bibitem[Iliakopoulou et~al\mbox{.}(2024)]%
        {llm-sched4}
\bibfield{author}{\bibinfo{person}{Nikoleta Iliakopoulou},
  \bibinfo{person}{Jovan Stojkovic}, \bibinfo{person}{Chloe Alverti},
  \bibinfo{person}{Tianyin Xu}, \bibinfo{person}{Hubertus Franke}, {and}
  \bibinfo{person}{Josep Torrellas}.} \bibinfo{year}{2024}\natexlab{}.
\newblock \bibinfo{title}{Chameleon: Adaptive Caching and Scheduling for
  Many-Adapter LLM Inference Environments}.
\newblock
\newblock
\showeprint[arxiv]{2411.17741}~[cs.DC]
\urldef\tempurl%
\url{https://arxiv.org/abs/2411.17741}
\showURL{%
\tempurl}


\bibitem[Jain et~al\mbox{.}(2025)]%
        {llm-sched3}
\bibfield{author}{\bibinfo{person}{Kunal Jain}, \bibinfo{person}{Anjaly
  Parayil}, \bibinfo{person}{Ankur Mallick}, \bibinfo{person}{Esha Choukse},
  \bibinfo{person}{Xiaoting Qin}, \bibinfo{person}{Jue Zhang},
  \bibinfo{person}{\'{I}\~{n}igo Goiri}, \bibinfo{person}{Rujia Wang},
  \bibinfo{person}{Chetan Bansal}, \bibinfo{person}{Victor R\"{u}hle},
  \bibinfo{person}{Anoop Kulkarni}, \bibinfo{person}{Steve Kofsky}, {and}
  \bibinfo{person}{Saravan Rajmohan}.} \bibinfo{year}{2025}\natexlab{}.
\newblock \showarticletitle{Performance Aware LLM Load Balancer for Mixed
  Workloads}. In \bibinfo{booktitle}{\emph{Proceedings of the 5th Workshop on
  Machine Learning and Systems}} (World Trade Center, Rotterdam, Netherlands)
  \emph{(\bibinfo{series}{EuroMLSys '25})}. \bibinfo{publisher}{Association for
  Computing Machinery}, \bibinfo{address}{New York, NY, USA},
  \bibinfo{pages}{19–30}.
\newblock
\showISBNx{9798400715389}
\urldef\tempurl%
\url{https://doi.org/10.1145/3721146.3721947}
\showDOI{\tempurl}


\bibitem[Kwon et~al\mbox{.}(2023)]%
        {noauthor_vllm-projectvllm_2024}
\bibfield{author}{\bibinfo{person}{Woosuk Kwon}, \bibinfo{person}{Zhuohan Li},
  \bibinfo{person}{Siyuan Zhuang}, \bibinfo{person}{Ying Sheng},
  \bibinfo{person}{Lianmin Zheng}, \bibinfo{person}{Cody~Hao Yu},
  \bibinfo{person}{Joseph~E. Gonzalez}, \bibinfo{person}{Hao Zhang}, {and}
  \bibinfo{person}{Ion Stoica}.} \bibinfo{year}{2023}\natexlab{}.
\newblock \showarticletitle{Efficient Memory Management for Large Language
  Model Serving with PagedAttention}. In \bibinfo{booktitle}{\emph{Proceedings
  of the ACM SIGOPS 29th Symposium on Operating Systems Principles}}.
\newblock


\bibitem[Lee et~al\mbox{.}(2024)]%
        {infinigen}
\bibfield{author}{\bibinfo{person}{Wonbeom Lee}, \bibinfo{person}{Jungi Lee},
  \bibinfo{person}{Junghwan Seo}, {and} \bibinfo{person}{Jaewoong Sim}.}
  \bibinfo{year}{2024}\natexlab{}.
\newblock \showarticletitle{{InfiniGen}: Efficient Generative Inference of
  Large Language Models with Dynamic {KV} Cache Management}. In
  \bibinfo{booktitle}{\emph{18th USENIX Symposium on Operating Systems Design
  and Implementation (OSDI 24)}}. \bibinfo{publisher}{USENIX Association},
  \bibinfo{address}{Santa Clara, CA}, \bibinfo{pages}{155--172}.
\newblock
\showISBNx{978-1-939133-40-3}
\urldef\tempurl%
\url{https://www.usenix.org/conference/osdi24/presentation/lee}
\showURL{%
\tempurl}


\bibitem[LeewayHertz(2025)]%
        {peft-compare}
\bibfield{author}{\bibinfo{person}{LeewayHertz}.}
  \bibinfo{year}{2025}\natexlab{}.
\newblock \bibinfo{title}{Parameter-efficient Fine-tuning (PEFT): Overview,
  benefits, techniques and model training}.
\newblock
\newblock
\urldef\tempurl%
\url{https://www.leewayhertz.com/parameter-efficient-fine-tuning/}
\showURL{%
\tempurl}


\bibitem[Lewis et~al\mbox{.}(2020)]%
        {rag}
\bibfield{author}{\bibinfo{person}{Patrick Lewis}, \bibinfo{person}{Ethan
  Perez}, \bibinfo{person}{Aleksandra Piktus}, \bibinfo{person}{Fabio Petroni},
  \bibinfo{person}{Vladimir Karpukhin}, \bibinfo{person}{Naman Goyal},
  \bibinfo{person}{Heinrich K{\"u}ttler}, \bibinfo{person}{Mike Lewis},
  \bibinfo{person}{Wen-tau Yih}, \bibinfo{person}{Tim Rockt{\"a}schel},
  {et~al\mbox{.}}} \bibinfo{year}{2020}\natexlab{}.
\newblock \showarticletitle{Retrieval-augmented generation for
  knowledge-intensive nlp tasks}.
\newblock \bibinfo{journal}{\emph{Advances in Neural Information Processing
  Systems}}  \bibinfo{volume}{33} (\bibinfo{year}{2020}),
  \bibinfo{pages}{9459--9474}.
\newblock


\bibitem[Li et~al\mbox{.}(2024)]%
        {li2024mixlora}
\bibfield{author}{\bibinfo{person}{Dengchun Li}, \bibinfo{person}{Yingzi Ma},
  \bibinfo{person}{Naizheng Wang}, \bibinfo{person}{Zhengmao Ye},
  \bibinfo{person}{Zhiyuan Cheng}, \bibinfo{person}{Yinghao Tang},
  \bibinfo{person}{Yan Zhang}, \bibinfo{person}{Lei Duan}, \bibinfo{person}{Jie
  Zuo}, \bibinfo{person}{Cal Yang}, {and} \bibinfo{person}{Mingjie Tang}.}
  \bibinfo{year}{2024}\natexlab{}.
\newblock \showarticletitle{MIXLORA: Enhancing Large Language Models
  Fine-Tuning with LoRA-based Mixture of Experts}.
\newblock \bibinfo{journal}{\emph{arXiv preprint arXiv:2404.15159v3}}
  (\bibinfo{year}{2024}).
\newblock


\bibitem[Li et~al\mbox{.}({[n.\,d.]})]%
        {noauthor_caraserve_nodate}
\bibfield{author}{\bibinfo{person}{Suyi Li}, \bibinfo{person}{Hanfeng Lu},
  \bibinfo{person}{Tianyuan Wu}, \bibinfo{person}{Minchen Yu},
  \bibinfo{person}{Qizhen Weng}, \bibinfo{person}{Xusheng Chen},
  \bibinfo{person}{Yizhou Shan}, \bibinfo{person}{Binhang Yuan}, {and}
  \bibinfo{person}{Wei Wang}.} \bibinfo{year}{[n.\,d.]}\natexlab{}.
\newblock \bibinfo{title}{{CaraServe}: {CPU}-{Assisted} and {Rank}-{Aware}
  {LoRA} {Serving} for {Generative} {LLM} {Inference}}.
\newblock
\newblock
\urldef\tempurl%
\url{https://arxiv.org/html/2401.11240v1}
\showURL{%
\tempurl}


\bibitem[Lialin et~al\mbox{.}(2024)]%
        {peft-comparison}
\bibfield{author}{\bibinfo{person}{Vladislav Lialin}, \bibinfo{person}{Vijeta
  Deshpande}, \bibinfo{person}{Xiaowei Yao}, {and} \bibinfo{person}{Anna
  Rumshisky}.} \bibinfo{year}{2024}\natexlab{}.
\newblock \showarticletitle{Scaling Down to Scale Up: A Guide to
  Parameter-Efficient Fine-Tuning}.
\newblock \bibinfo{journal}{\emph{arXiv preprint arXiv:2303.15647}}
  (\bibinfo{year}{2024}).
\newblock


\bibitem[Liang et~al\mbox{.}(2023)]%
        {me-attack}
\bibfield{author}{\bibinfo{person}{Jiacheng Liang}, \bibinfo{person}{Ren Pang},
  \bibinfo{person}{Changjiang Li}, {and} \bibinfo{person}{Ting Wang}.}
  \bibinfo{year}{2023}\natexlab{}.
\newblock \showarticletitle{Model Extraction Attacks Revisited}.
\newblock \bibinfo{journal}{\emph{arXiv preprint arXiv:2312.05386}}
  (\bibinfo{year}{2023}).
\newblock


\bibitem[Liu et~al\mbox{.}(2022)]%
        {ia3}
\bibfield{author}{\bibinfo{person}{Haokun Liu}, \bibinfo{person}{Derek Tam},
  \bibinfo{person}{Mohammed Muqeeth}, \bibinfo{person}{Jay Mohta},
  \bibinfo{person}{Tenghao Huang}, \bibinfo{person}{Mohit Bansal}, {and}
  \bibinfo{person}{Colin Raffel}.} \bibinfo{year}{2022}\natexlab{}.
\newblock \showarticletitle{Few-Shot Parameter-Efficient Fine-Tuning is Better
  and Cheaper than In-Context Learning}.
\newblock \bibinfo{journal}{\emph{arXiv preprint arXiv:2205.05638}}
  (\bibinfo{year}{2022}).
\newblock


\bibitem[Mangrulkar et~al\mbox{.}(2022)]%
        {peft}
\bibfield{author}{\bibinfo{person}{Sourab Mangrulkar}, \bibinfo{person}{Sylvain
  Gugger}, \bibinfo{person}{Lysandre Debut}, \bibinfo{person}{Younes Belkada},
  \bibinfo{person}{Sayak Paul}, {and} \bibinfo{person}{Benjamin Bossan}.}
  \bibinfo{year}{2022}\natexlab{}.
\newblock \bibinfo{title}{PEFT: State-of-the-art Parameter-Efficient
  Fine-Tuning methods}.
\newblock \bibinfo{howpublished}{\url{https://github.com/huggingface/peft}}.
\newblock


\bibitem[Nikdan et~al\mbox{.}(2024)]%
        {rosa}
\bibfield{author}{\bibinfo{person}{Mahdi Nikdan}, \bibinfo{person}{Soroush
  Tabesh}, \bibinfo{person}{Elvir Crncevic}, {and} \bibinfo{person}{Dan
  Alistarh}.} \bibinfo{year}{2024}\natexlab{}.
\newblock \showarticletitle{RoSA: Accurate Parameter-Efficient Fine-Tuning via
  Robust Adaptation}.
\newblock \bibinfo{journal}{\emph{arXiv preprint arXiv:2401.04679}}
  (\bibinfo{year}{2024}).
\newblock
\urldef\tempurl%
\url{https://arxiv.org/abs/2401.04679}
\showURL{%
\tempurl}


\bibitem[Oliaro et~al\mbox{.}(2025)]%
        {flexllm}
\bibfield{author}{\bibinfo{person}{Gabriele Oliaro}, \bibinfo{person}{Xupeng
  Miao}, \bibinfo{person}{Xinhao Cheng}, \bibinfo{person}{Vineeth Kada},
  \bibinfo{person}{Ruohan Gao}, \bibinfo{person}{Yingyi Huang},
  \bibinfo{person}{Remi Delacourt}, \bibinfo{person}{April Yang},
  \bibinfo{person}{Yingcheng Wang}, \bibinfo{person}{Mengdi Wu},
  \bibinfo{person}{Colin Unger}, {and} \bibinfo{person}{Zhihao Jia}.}
  \bibinfo{year}{2025}\natexlab{}.
\newblock \bibinfo{title}{FlexLLM: A System for Co-Serving Large Language Model
  Inference and Parameter-Efficient Finetuning}.
\newblock
\newblock
\showeprint[arxiv]{2402.18789}~[cs.DC]
\urldef\tempurl%
\url{https://arxiv.org/abs/2402.18789}
\showURL{%
\tempurl}


\bibitem[Pan et~al\mbox{.}(2024)]%
        {pan2024instinfer}
\bibfield{author}{\bibinfo{person}{Xiurui Pan}, \bibinfo{person}{Endian Li},
  \bibinfo{person}{Qiao Li}, \bibinfo{person}{Shengwen Liang},
  \bibinfo{person}{Yizhou Shan}, \bibinfo{person}{Ke Zhou},
  \bibinfo{person}{Yingwei Luo}, \bibinfo{person}{Xiaolin Wang}, {and}
  \bibinfo{person}{Jie Zhang}.} \bibinfo{year}{2024}\natexlab{}.
\newblock \showarticletitle{InstInfer: In-Storage Attention Offloading for
  Cost-Effective Long-Context LLM Inference}.
\newblock \bibinfo{journal}{\emph{arXiv preprint arXiv:2409.04992}}
  (\bibinfo{year}{2024}).
\newblock


\bibitem[Patel et~al\mbox{.}(2024)]%
        {splitwise}
\bibfield{author}{\bibinfo{person}{Pratyush Patel}, \bibinfo{person}{Esha
  Choukse}, \bibinfo{person}{Chaojie Zhang}, \bibinfo{person}{Aashaka Shah},
  \bibinfo{person}{Íñigo Goiri}, \bibinfo{person}{Saeed Maleki}, {and}
  \bibinfo{person}{Ricardo Bianchini}.} \bibinfo{year}{2024}\natexlab{}.
\newblock \showarticletitle{Splitwise: Efficient generative LLM inference using
  phase splitting}. In \bibinfo{booktitle}{\emph{ISCA}}.
\newblock
\urldef\tempurl%
\url{https://www.microsoft.com/en-us/research/publication/splitwise-efficient-generative-llm-inference-using-phase-splitting/}
\showURL{%
\tempurl}


\bibitem[Qin et~al\mbox{.}(2025)]%
        {mooncake}
\bibfield{author}{\bibinfo{person}{Ruoyu Qin}, \bibinfo{person}{Zheming Li},
  \bibinfo{person}{Weiran He}, \bibinfo{person}{Jialei Cui},
  \bibinfo{person}{Feng Ren}, \bibinfo{person}{Mingxing Zhang},
  \bibinfo{person}{Yongwei Wu}, \bibinfo{person}{Weimin Zheng}, {and}
  \bibinfo{person}{Xinran Xu}.} \bibinfo{year}{2025}\natexlab{}.
\newblock \showarticletitle{Mooncake: Trading More Storage for Less Computation
  {\textemdash} A {KVCache-centric} Architecture for Serving {LLM} Chatbot}. In
  \bibinfo{booktitle}{\emph{23rd USENIX Conference on File and Storage
  Technologies (FAST 25)}}. \bibinfo{publisher}{USENIX Association},
  \bibinfo{address}{Santa Clara, CA}, \bibinfo{pages}{155--170}.
\newblock
\showISBNx{978-1-939133-45-8}
\urldef\tempurl%
\url{https://www.usenix.org/conference/fast25/presentation/qin}
\showURL{%
\tempurl}


\bibitem[Rajbhandari et~al\mbox{.}(2022)]%
        {deepspeed2}
\bibfield{author}{\bibinfo{person}{Samyam Rajbhandari},
  \bibinfo{person}{Conglong Li}, \bibinfo{person}{Zhewei Yao},
  \bibinfo{person}{Minjia Zhang}, \bibinfo{person}{Reza~Yazdani Aminabadi},
  \bibinfo{person}{Ammar~Ahmad Awan}, \bibinfo{person}{Jeff Rasley}, {and}
  \bibinfo{person}{Yuxiong He}.} \bibinfo{year}{2022}\natexlab{}.
\newblock \showarticletitle{Deepspeed-moe: Advancing mixture-of-experts
  inference and training to power next-generation ai scale}. In
  \bibinfo{booktitle}{\emph{Proceedings of the International Conference on
  Machine Learning (ICML)}}.
\newblock


\bibitem[Rajbhandari et~al\mbox{.}(2021)]%
        {zeroinfinity}
\bibfield{author}{\bibinfo{person}{Samyam Rajbhandari},
  \bibinfo{person}{Olatunji Ruwase}, \bibinfo{person}{Jeff Rasley},
  \bibinfo{person}{Shaden Smith}, {and} \bibinfo{person}{Yuxiong He}.}
  \bibinfo{year}{2021}\natexlab{}.
\newblock \showarticletitle{Zero-infinity: Breaking the gpu memory wall for
  extreme scale deep learning}. In \bibinfo{booktitle}{\emph{Proceedings of the
  International Conference for High Performance Computing, Networking, Storage
  and Analysis (SC)}}.
\newblock


\bibitem[Sheng et~al\mbox{.}(2024)]%
        {sheng_s-lora_2024}
\bibfield{author}{\bibinfo{person}{Ying Sheng}, \bibinfo{person}{Shiyi Cao},
  \bibinfo{person}{Dacheng Li}, \bibinfo{person}{Coleman Hooper},
  \bibinfo{person}{Nicholas Lee}, \bibinfo{person}{Shuo Yang},
  \bibinfo{person}{Christopher Chou}, \bibinfo{person}{Banghua Zhu},
  \bibinfo{person}{Lianmin Zheng}, \bibinfo{person}{Kurt Keutzer},
  \bibinfo{person}{Joseph~E. Gonzalez}, {and} \bibinfo{person}{Ion Stoica}.}
  \bibinfo{year}{2024}\natexlab{}.
\newblock \bibinfo{title}{S-{LoRA}: {Serving} {Thousands} of {Concurrent}
  {LoRA} {Adapters}}.
\newblock
\newblock
\urldef\tempurl%
\url{https://doi.org/10.48550/arXiv.2311.03285}
\showDOI{\tempurl}
\newblock
\shownote{arXiv:2311.03285 [cs]}.


\bibitem[Sheng et~al\mbox{.}(2023)]%
        {flexgen}
\bibfield{author}{\bibinfo{person}{Ying Sheng}, \bibinfo{person}{Lianmin
  Zheng}, \bibinfo{person}{Binhang Yuan}, \bibinfo{person}{Zhuohan Li},
  \bibinfo{person}{Max Ryabinin}, \bibinfo{person}{Daniel~Y. Fu},
  \bibinfo{person}{Zhiqiang Xie}, \bibinfo{person}{Beidi Chen},
  \bibinfo{person}{Clark Barrett}, \bibinfo{person}{Joseph~E. Gonzalez},
  \bibinfo{person}{Percy Liang}, \bibinfo{person}{Christopher Ré},
  \bibinfo{person}{Ion Stoica}, {and} \bibinfo{person}{Ce Zhang}.}
  \bibinfo{year}{2023}\natexlab{}.
\newblock \bibinfo{title}{{FlexGen: High-Throughput Generative Inference of
  Large Language Models with a Single GPU}}.
\newblock
\newblock
\urldef\tempurl%
\url{https://arxiv.org/abs/2303.06865}
\showURL{%
\tempurl}


\bibitem[Tram{\`e}r et~al\mbox{.}(2016)]%
        {me-attack3}
\bibfield{author}{\bibinfo{person}{Florian Tram{\`e}r}, \bibinfo{person}{Fan
  Zhang}, \bibinfo{person}{Ari Juels}, \bibinfo{person}{Michael~K. Reiter},
  {and} \bibinfo{person}{Thomas Ristenpart}.} \bibinfo{year}{2016}\natexlab{}.
\newblock \showarticletitle{Stealing Machine Learning Models via Prediction
  {APIs}}. In \bibinfo{booktitle}{\emph{25th USENIX Security Symposium (USENIX
  Security 16)}}. \bibinfo{publisher}{USENIX Association},
  \bibinfo{address}{Austin, TX}, \bibinfo{pages}{601--618}.
\newblock
\showISBNx{978-1-931971-32-4}
\urldef\tempurl%
\url{https://www.usenix.org/conference/usenixsecurity16/technical-sessions/presentation/tramer}
\showURL{%
\tempurl}


\bibitem[Wang et~al\mbox{.}(2025)]%
        {honeypotnet}
\bibfield{author}{\bibinfo{person}{Yixu Wang}, \bibinfo{person}{Tianle Gu},
  \bibinfo{person}{Yan Teng}, \bibinfo{person}{Yingchun Wang}, {and}
  \bibinfo{person}{Xingjun Ma}.} \bibinfo{year}{2025}\natexlab{}.
\newblock \showarticletitle{HoneypotNet: Backdoor Attacks Against Model
  Extraction}.
\newblock \bibinfo{journal}{\emph{arXiv preprint arXiv:2501.01090}}
  (\bibinfo{year}{2025}).
\newblock


\bibitem[Wolf et~al\mbox{.}(2020a)]%
        {transformers}
\bibfield{author}{\bibinfo{person}{Thomas Wolf}, \bibinfo{person}{Lysandre
  Debut}, \bibinfo{person}{Victor Sanh}, \bibinfo{person}{Julien Chaumond},
  \bibinfo{person}{Clement Delangue}, \bibinfo{person}{Anthony Moi},
  \bibinfo{person}{Perric Cistac}, \bibinfo{person}{Clara Ma},
  \bibinfo{person}{Yacine Jernite}, \bibinfo{person}{Julien Plu},
  \bibinfo{person}{Canwen Xu}, \bibinfo{person}{Teven Le~Scao},
  \bibinfo{person}{Sylvain Gugger}, \bibinfo{person}{Mariama Drame},
  \bibinfo{person}{Quentin Lhoest}, {and} \bibinfo{person}{Alexander~M. Rush}.}
  \bibinfo{year}{2020}\natexlab{a}.
\newblock \showarticletitle{{Transformers: State-of-the-Art Natural Language
  Processing}}. \bibinfo{publisher}{Association for Computational Linguistics},
  \bibinfo{pages}{38--45}.
\newblock
\urldef\tempurl%
\url{https://www.aclweb.org/anthology/2020.emnlp-demos.6}
\showURL{%
\tempurl}


\bibitem[Wolf et~al\mbox{.}(2020b)]%
        {tx}
\bibfield{author}{\bibinfo{person}{Thomas Wolf}, \bibinfo{person}{Lysandre
  Debut}, \bibinfo{person}{Victor Sanh}, \bibinfo{person}{Julien Chaumond},
  \bibinfo{person}{Clement Delangue}, \bibinfo{person}{Anthony Moi},
  \bibinfo{person}{Pierric Cistac}, \bibinfo{person}{Tim Rault},
  \bibinfo{person}{Rémi Louf}, \bibinfo{person}{Morgan Funtowicz},
  \bibinfo{person}{Joe Davison}, \bibinfo{person}{Sam Shleifer},
  \bibinfo{person}{Patrick von Platen}, \bibinfo{person}{Clara Ma},
  \bibinfo{person}{Yacine Jernite}, \bibinfo{person}{Julien Plu},
  \bibinfo{person}{Canwen Xu}, \bibinfo{person}{Teven~Le Scao},
  \bibinfo{person}{Sylvain Gugger}, \bibinfo{person}{Mariama Drame},
  \bibinfo{person}{Quentin Lhoest}, {and} \bibinfo{person}{Alexander~M. Rush}.}
  \bibinfo{year}{2020}\natexlab{b}.
\newblock \showarticletitle{Transformers: State-of-the-Art Natural Language
  Processing}. In \bibinfo{booktitle}{\emph{Proceedings of the 2020 Conference
  on Empirical Methods in Natural Language Processing: System Demonstrations}}.
  \bibinfo{publisher}{Association for Computational Linguistics},
  \bibinfo{address}{Online}, \bibinfo{pages}{38--45}.
\newblock
\urldef\tempurl%
\url{https://www.aclweb.org/anthology/2020.emnlp-demos.6}
\showURL{%
\tempurl}


\bibitem[Wu et~al\mbox{.}(2024a)]%
        {llm-sched2}
\bibfield{author}{\bibinfo{person}{Bingyang Wu}, \bibinfo{person}{Yinmin
  Zhong}, \bibinfo{person}{Zili Zhang}, \bibinfo{person}{Shengyu Liu},
  \bibinfo{person}{Fangyue Liu}, \bibinfo{person}{Yuanhang Sun},
  \bibinfo{person}{Gang Huang}, \bibinfo{person}{Xuanzhe Liu}, {and}
  \bibinfo{person}{Xin Jin}.} \bibinfo{year}{2024}\natexlab{a}.
\newblock \bibinfo{title}{Fast Distributed Inference Serving for Large Language
  Models}.
\newblock
\newblock
\showeprint[arxiv]{2305.05920}~[cs.LG]
\urldef\tempurl%
\url{https://arxiv.org/abs/2305.05920}
\showURL{%
\tempurl}


\bibitem[Wu et~al\mbox{.}(2024b)]%
        {dlora}
\bibfield{author}{\bibinfo{person}{Bingyang Wu}, \bibinfo{person}{Ruidong Zhu},
  \bibinfo{person}{Zili Zhang}, \bibinfo{person}{Peng Sun},
  \bibinfo{person}{Xuanzhe Liu}, {and} \bibinfo{person}{Xin Jin}.}
  \bibinfo{year}{2024}\natexlab{b}.
\newblock \showarticletitle{{dLoRA}: Dynamically Orchestrating Requests and
  Adapters for {LoRA} {LLM} Serving}. In \bibinfo{booktitle}{\emph{18th USENIX
  Symposium on Operating Systems Design and Implementation (OSDI 24)}}.
  \bibinfo{publisher}{USENIX Association}, \bibinfo{address}{Santa Clara, CA},
  \bibinfo{pages}{911--927}.
\newblock
\showISBNx{978-1-939133-40-3}
\urldef\tempurl%
\url{https://www.usenix.org/conference/osdi24/presentation/wu-bingyang}
\showURL{%
\tempurl}


\bibitem[Xiao et~al\mbox{.}(2024)]%
        {duo-attention}
\bibfield{author}{\bibinfo{person}{Guangxuan Xiao}, \bibinfo{person}{Jiaming
  Tang}, \bibinfo{person}{Jingwei Zuo}, \bibinfo{person}{Junxian Guo},
  \bibinfo{person}{Shang Yang}, \bibinfo{person}{Haotian Tang},
  \bibinfo{person}{Yao Fu}, {and} \bibinfo{person}{Song Han}.}
  \bibinfo{year}{2024}\natexlab{}.
\newblock \bibinfo{title}{DuoAttention: Efficient Long-Context LLM Inference
  with Retrieval and Streaming Heads}.
\newblock
\newblock
\showeprint[arxiv]{2410.10819}~[cs.CL]
\urldef\tempurl%
\url{https://arxiv.org/abs/2410.10819}
\showURL{%
\tempurl}


\bibitem[Xiong et~al\mbox{.}(2024)]%
        {layerkv}
\bibfield{author}{\bibinfo{person}{Yi Xiong}, \bibinfo{person}{Hao Wu},
  \bibinfo{person}{Changxu Shao}, \bibinfo{person}{Ziqing Wang},
  \bibinfo{person}{Rui Zhang}, \bibinfo{person}{Yuhong Guo},
  \bibinfo{person}{Junping Zhao}, \bibinfo{person}{Ke Zhang}, {and}
  \bibinfo{person}{Zhenxuan Pan}.} \bibinfo{year}{2024}\natexlab{}.
\newblock \showarticletitle{LayerKV: Optimizing Large Language Model Serving
  with Layer-wise KV Cache Management}.
\newblock \bibinfo{journal}{\emph{arXiv preprint arXiv:2410.00428v3}}
  (\bibinfo{year}{2024}).
\newblock


\bibitem[Ye et~al\mbox{.}(2023)]%
        {aspen}
\bibfield{author}{\bibinfo{person}{Zhengmao Ye}, \bibinfo{person}{Dengchun Li},
  \bibinfo{person}{Jingqi Tian}, \bibinfo{person}{Tingfeng Lan},
  \bibinfo{person}{Jie Zuo}, \bibinfo{person}{Lei Duan}, \bibinfo{person}{Hui
  Lu}, \bibinfo{person}{Yexi Jiang}, \bibinfo{person}{Jian Sha},
  \bibinfo{person}{Ke Zhang}, {and} \bibinfo{person}{Mingjie Tang}.}
  \bibinfo{year}{2023}\natexlab{}.
\newblock \showarticletitle{ASPEN: High-Throughput LoRA Fine-Tuning of Large
  Language Models with a Single GPU}.
\newblock \bibinfo{journal}{\emph{arXiv preprint arXiv:2312.02515}}
  (\bibinfo{year}{2023}).
\newblock


\bibitem[Yu et~al\mbox{.}(2022)]%
        {orca}
\bibfield{author}{\bibinfo{person}{Gyeong-In Yu}, \bibinfo{person}{Joo~Seong
  Jeong}, \bibinfo{person}{Geon-Woo Kim}, \bibinfo{person}{Soojeong Kim}, {and}
  \bibinfo{person}{Byung-Gon Chun}.} \bibinfo{year}{2022}\natexlab{}.
\newblock \showarticletitle{Orca: A Distributed Serving System for
  {Transformer-Based} Generative Models}. In \bibinfo{booktitle}{\emph{16th
  USENIX Symposium on Operating Systems Design and Implementation (OSDI 22)}}.
  \bibinfo{publisher}{USENIX Association}, \bibinfo{address}{Carlsbad, CA},
  \bibinfo{pages}{521--538}.
\newblock
\showISBNx{978-1-939133-28-1}
\urldef\tempurl%
\url{https://www.usenix.org/conference/osdi22/presentation/yu}
\showURL{%
\tempurl}


\bibitem[ZeroMQ(2024)]%
        {zeromq}
\bibfield{author}{\bibinfo{person}{ZeroMQ}.} \bibinfo{year}{2024}\natexlab{}.
\newblock \bibinfo{title}{Getting Started with ZeroMQ}.
\newblock
\newblock
\urldef\tempurl%
\url{https://zeromq.org/get-started/}
\showURL{%
\tempurl}
\newblock
\shownote{Accessed: 2024-09-23}.


\bibitem[Zhang et~al\mbox{.}(2023)]%
        {h2o}
\bibfield{author}{\bibinfo{person}{Zhenyu Zhang}, \bibinfo{person}{Ying Sheng},
  \bibinfo{person}{Tianyi Zhou}, \bibinfo{person}{Tianlong Chen},
  \bibinfo{person}{Lianmin Zheng}, \bibinfo{person}{Ruisi Cai},
  \bibinfo{person}{Zhao Song}, \bibinfo{person}{Yuandong Tian},
  \bibinfo{person}{Christopher Re}, \bibinfo{person}{Clark Barrett},
  \bibinfo{person}{Zhangyang Wang}, {and} \bibinfo{person}{Beidi Chen}.}
  \bibinfo{year}{2023}\natexlab{}.
\newblock \showarticletitle{H2O: Heavy-Hitter Oracle for Efficient Generative
  Inference of Large Language Models}. In
  \bibinfo{booktitle}{\emph{Thirty-seventh Conference on Neural Information
  Processing Systems}}.
\newblock
\urldef\tempurl%
\url{https://openreview.net/forum?id=RkRrPp7GKO}
\showURL{%
\tempurl}


\bibitem[Zhengmao et~al\mbox{.}(2023)]%
        {mLoRA2023}
\bibfield{author}{\bibinfo{person}{Ye Zhengmao}, \bibinfo{person}{Li Dengchun},
  \bibinfo{person}{Tian Jingqi}, \bibinfo{person}{Lan Tingfeng},
  \bibinfo{person}{Liang Yanbo}, \bibinfo{person}{Jiang Yexi},
  \bibinfo{person}{Zuo Jie}, \bibinfo{person}{Lu Hui}, \bibinfo{person}{Duan
  Lei}, {and} \bibinfo{person}{Tang Mingjie}.} \bibinfo{year}{2023}\natexlab{}.
\newblock \bibinfo{title}{m-LoRA: Efficient LLM Model Fine-tune and Inference
  via Multi-Lora Optimization}.
\newblock \bibinfo{howpublished}{\url{https://github.com/TUDB-Labs/mLoRA}}.
\newblock
\newblock
\shownote{*: these authors contributed equally to this work.}.


\bibitem[Zhong et~al\mbox{.}(2024)]%
        {distserve}
\bibfield{author}{\bibinfo{person}{Yinmin Zhong}, \bibinfo{person}{Shengyu
  Liu}, \bibinfo{person}{Junda Chen}, \bibinfo{person}{Jianbo Hu},
  \bibinfo{person}{Yibo Zhu}, \bibinfo{person}{Xuanzhe Liu},
  \bibinfo{person}{Xin Jin}, {and} \bibinfo{person}{Hao Zhang}.}
  \bibinfo{year}{2024}\natexlab{}.
\newblock \bibinfo{title}{DistServe: Disaggregating Prefill and Decoding for
  Goodput-optimized Large Language Model Serving}.
\newblock
\newblock
\showeprint[arxiv]{2401.09670}~[cs.DC]
\urldef\tempurl%
\url{https://arxiv.org/abs/2401.09670}
\showURL{%
\tempurl}


\end{thebibliography}
\bibliographystyle{ACM-Reference-Format}
\end{document}